\documentclass[a4paper,fleqn,usenatbib]{mnras}

\usepackage{newtxtext,newtxmath}

\usepackage[T1]{fontenc}
\usepackage{ae,aecompl}

\usepackage{amssymb}
\usepackage{graphicx}
\usepackage{amsmath}
\usepackage{bm}

\usepackage{alltt}  
\usepackage{graphicx, subfigure}
\usepackage{multirow}
\usepackage[table]{xcolor}  
\usepackage{pbox}
\usepackage{float}
\usepackage{afterpage}
\usepackage{longtable}
\usepackage{color}
\definecolor{darkgreen}{rgb}{0.0, 0.4, 0.0}
\definecolor{darkblue}{rgb}{0.0, 0.0, 0.6}
\definecolor{darkred}{rgb}{0.6, 0.0, 0.0}

\AtBeginShipout{%
  \ifnum\value{page}>1 %
    \typeout{* Additional boxing of page `\thepage'}%
    \setbox\AtBeginShipoutBox=\hbox{\copy\AtBeginShipoutBox}%
  \fi
}

\usepackage{lscape}

\begin{document}
\title[Exomoon Habitability and Tidal Evolution]{Exomoon Habitability and Tidal Evolution in Low-Mass Star Systems}


\author[R. R. Zollinger et al.]{
Rhett R. Zollinger,$^{1}$\thanks{E-mail: rhettzollinger@suu.edu}
John C. Armstrong,$^{2}$\thanks{E-mail: jcarmstrong@weber.edu}
Ren$\acute{\mathrm{e}}$ Heller$^{3}$\thanks{E-mail: heller@mps.mpg.de}
\\
$^{1}$Southern Utah University, Cedar City, Utah 84720, USA\\
$^{2}$Weber State University, Ogden, Utah 84408, USA\\
$^{3}$Max Planck Institute for Solar System Research, Justus-von-Liebig-Weg 3, 37077 G\"ottingen, Germany
}




\date{Accepted 2017 July 20. Received 2017 July 14; in original form 2016 August 9}

\pubyear{2017}

\label{firstpage}
\pagerange{\pageref{firstpage}--\pageref{lastpage}}
\maketitle

\begin{abstract}
Discoveries of extrasolar planets in the habitable zone (HZ) of their parent star lead to questions about the habitability of massive moons orbiting planets in the HZ. Around \textcolor{black}{low-mass} stars, the HZ is much closer to the star than for Sun-like stars. For a planet-moon binary in such a HZ, the proximity of the star forces a close orbit for the moon to remain gravitationally bound to the planet. Under these conditions the effects of tidal heating, distortion torques, and stellar perturbations become important considerations for exomoon habitability. 

Utilizing a model that considers both dynamical and tidal interactions simultaneously, we performed a computational investigation into exomoon evolution for systems in the HZ of \textcolor{black}{low-mass stars ($\lesssim 0.6\ M_{\odot}$)}. We show that dwarf stars with masses $\lesssim 0.2\ M_{\odot}$ cannot host habitable exomoons within the stellar HZ due to extreme tidal heating in the moon. Perturbations from a central star may continue to have deleterious effects in the HZ up to $\approx 0.5\ M_{\odot}$, depending on the host planet's mass and its location in the HZ, amongst others. In addition to heating concerns, torques due to tidal and spin distortion can lead to the relatively rapid inward spiraling of a moon. Therefore, moons of giant planets in HZs around the most abundant type of star are unlikely to have habitable surfaces. In cases with lower intensity tidal heating the stellar perturbations may have a positive influence on exomoon habitability by promoting long-term heating and possibly extending the HZ for exomoons.
\end{abstract}




\section{Introduction} \label{intro}

The exploration of the moons of Jupiter and Saturn has provided immense understanding of otherworldly environments. Some of these moons have reservoirs of liquids, nutrients, and internal heat \citep{1983Natur.301..225S,2006Sci...311.1422H,2008Natur.454..607B,2015JGRA..120.1715S} -- three basic components that are essential for life on Earth. Recent technological and theoretical advances in astronomy and biology now raise the question of whether life might exist on any of the moons beyond the solar system (``exomoons''). Given the abundant population of moons in our system, exomoons may be even more numerous than exoplanets \citep{2015ApJ...806..181H}.

No moon outside the Solar System has been detected, but the first detection of an extrasolar moon appears to be on the horizon \citep{kipping09,kipping12,heller14} now that modern techniques enable detections of sub-Earth sized extrasolar planets \citep{2012ApJ...747..144M,2013Natur.494..452B}. A recent review of current theories on the formation, detection, and habitability of exomoons suggests that natural satellites in the range of 0.1-0.5 Earth masses (i) are potentially habitable, (ii) can form within the circumplanetary debris and gas disk or via capture from a binary, and (iii) are detectable with current technology \citep{heller14}. \textcolor{black}{Considering the potential for current observation}, we explore the expected properties of such exomoons in this paper.

When investigating planet habitability, the primary heat source is typically the radiated energy from a central star. Considerations of stellar radiation and planet surface temperatures have led to the adoption of a concept referred to as the stellar ``habitable zone'' (HZ) that is the region around a star in which an Earth-like planet with an Earth-like atmosphere can sustain liquid surface water \citep{kasting1993}. Tidal heating is typically less important as an energy source for planets in the HZ, though it might be relevant for eccentric planets in the HZs around \textcolor{black}{low-mass} stars of spectral type M \citep{2008AsBio...8..557B}. For the moon of a large planet, tidal heating can work as an alternative internal heat source as well. Several studies have addressed the importance of tidal heating and its effects on the habitability of exomoons \citep{reynolds1987,scharf06,henning09,2011ApJ...736L..14P,heller12,2013MNRAS.432.2994F,hb13,hz13,2015IJAsB..14..335H,2015ApJ...804...41D,2017arXiv170302447D}.  Tidal heat can potentially maintain internal heating over several Gyr \citep[see Jupiter's moon Io;][]{2000Sci...288.1198S}, which contributes to surface heating and potentially drives important internal processes such as plate tectonics. 

Tidal interactions between a massive moon and its host planet become particularly interesting for planets in the HZs of M dwarfs, which are smaller, cooler, fainter, and less massive than the Sun. They are the predominant stellar population of our Galaxy \citep{chabrier00}, and as such, these low-mass stars may be the most abundant planet hosts in our Galaxy \citep{2013PNAS..11019273P,2015ApJ...807...45D}. Their lower core temperatures and decreased energy output result in a HZ that is much closer to the star in comparison to Sun-like stars. If a planet in the HZ around an M dwarf star has a massive moon, the close-in orbital distances could potentially influence the orbit of the moon around the planet \citep{heller12}. The relatively close central star can serve as a continual source of gravitational perturbation to the moon's orbit, which has important implications to tidal heating and orbital evolution of moons in these systems.

\textcolor{black}{With the possible abundance of planetary systems around low-mass stars and their unique characteristics for habitability, we are interested in exploring their potential for exomoon habitability. To accomplish this we employed a distinctive approach to simultaneously consider both orbital and tidal influences. In this paper, we review some theories on tidal interactions between two massive bodies and then conduct a computational investigation into the importance of these interactions on the long-term evolution of hypothetical exomoons. For our study we focus on massive moons around large planets in the HZ of \textcolor{black}{low-mass stars ($\lesssim 0.6\ M_{\odot}$)} which demonstrates the long-term effect of gravitational perturbations from a central star.}

\section{Dynamical and Tidal Evolution}\label{dtevol}

Tidal bulges raised in both a planet and satellite will dissipate energy and apply torques between the two bodies. The rate of dissipation strongly depends on the distance between the two objects. As a result of the tidal drag from the planet, a massive satellite will be subject to essentially four effects on its spin-orbital configuration:

\begin{enumerate}
\item {\bf Semi-major axis:} Tidal torques can cause a moon to either spiral in or out \citep{barnes02,2012ApJ...754...51S}. The direction of the spiral depends on the alignment between the planet's tidal bulge (raised by the moon) and the line connecting the two centers of mass. If the planet's rotation period is shorter than the orbital period of the satellite, the bulge will lead (assuming prograde orbits) and the moon will slowly spiral outward. This action could eventually destabilize the moon's orbit, leading to its ejection. On the other hand, if the planet's rotation period is longer, the bulge will lag and the moon will slowly spiral inward. As this happens, the tidal forces on the moon become increasingly greater. If the inward migration continues past the Roche limit, the satellite can be disintegrated.
\item {\bf Eccentricity:} Non-circular orbits will be circularized over the longterm. The timescale for which the eccentricity is damped can be estimated as \citep{gs66}
\begin{equation} \label{tdamp} \tau_e \approx \frac {4}{63} \frac{M_{\rm s}}{M_{\rm p}} {\left(\frac {a}{R_{\rm s}}\right)}^5 \frac{Q'_{\rm s}}{n}\ , \end{equation} 
where $n$ is the orbital mean motion frequency, $Q'_{\rm s}$ is the tidal dissipation function, \textcolor{black}{$a$ is the semi-major axis, $M_{\rm s}$ and $R_{\rm s}$ are the mass and radius of the satellite, and $M_{\rm p}$ the mass of the host planet.}
\item {\bf Rotation frequency:} The moon will have its rotation frequency braked and ultimately synchronized with its orbital motion around the planet \citep{dole1964,porter11,2012ApJ...754...51S}, a state that is commonly known as tidal locking.
\item {\bf Rotation axis:} Any initial spin-orbit misalignment (or obliquity) will be eroded, causing the moon's rotation axis to be perpendicular to its orbit about the planet. In addition, a moon will inevitably orbit in the equatorial plane of the planet due to both the Kozai mechanism and tidal evolution \citep{porter11}. The combination of all these effects will result in the satellite having the same obliquity as the planet with respect to the circumstellar orbit. As for the host planet, massive planets are more likely to maintain their primordial spin-orbit misalignment than small planets \citep{heller11}. Therefore, satellites of giant planets are more likely to maintain an orbital tilt relative to the star than even a single terrestrial planet at the same distance from a star.
\end{enumerate}

The issue of tidal equilibrium in three-body hierarchical star-planet-moon systems in the Keplerian limit has recently been treated by \citet{2016arXiv160708170A}, who demonstrated that a moon in tidal equilibrium will have its circumplanetary orbit widened until it ultimately gets ejected by stellar perturbations. $N$-body gravitational perturbations from the star set an upper limit on the maximum orbital radius of the moon around its planet at about half the planetary Hill radius for prograde moons \citep{2006MNRAS.373.1227D}. However, moons can be ejected from their planet even from close orbits around their planets, e.g. when the planet-moon system migrates towards the star and stellar perturbations increase the moon's eccentricity fatally \citep{2016ApJ...817...18S}. Perturbations from other planets can also destruct exomoon systems during planet-planet encounters \citep{2013ApJ...769L..14G,2013ApJ...775L..44P,2015MNRAS.449..828H}.

\subsection{Tidal Heating} \label{tid_theory}

As a result of the tidal orbital evolution, orbital energy is transformed into heat in either the planet or the moon or both. Several quantitative models for tidal dissipation have been proposed, the most widely used family of which is commonly referred to as equilibrium tide models \citep{1879_Darwin,hut1981,efroimsky07,ferraz08,hansen10}. However, the exact mechanisms of tidal dissipation invoke feedback processes between the tidal response of the body and the material it consists of \citep{henning09,2014ApJ...789...30H}, which are neglected in most equilibrium tide models. A conventional model that assumes a constant phase lag between the tidal bulge and the line between the centers of mass quantifies the tidal heating ($H$) of a satellite as
\begin{equation} \label{eq:H} H = \frac{63}{4} \frac{(G M_{\rm p})^{3/2} M_{\rm p} R_{\rm s}^5}{Q'_{\rm s}} a^{-15/2} e^2, \end{equation}
where $G$ is the gravitational constant and $e$ is the satellite's eccentricity around the planet \citep{1978Icar...36..245P,jackson08a}. Equation~(\ref{eq:H}) implies that tidal heating drops off quickly with increasing distance and altogether ceases for circular orbits ($e$ = 0).\footnote{This model breaks down for large $e$ \citep{2009ApJ...698L..42G} and it neglects effects of tidal heating from obliquity erosion \citep{heller11} and from rotational synchronization.}

In Eqs.~(\ref{tdamp}) and (\ref{eq:H}), $Q'$ parameterizes the physical response of a body to tides \citep{peale1979}. Its specific usefulness is that it encapsulates all the uncertainties about the tidal dissipation mechanisms. \textcolor{black}{For solid bodies, this function can be related to the rigidity $\mu$ and the standard $Q$-value}, $Q' = Q(1 + 19{\mu}/2g{\rho}R)$, where $g$ is the gravitational acceleration at the surface of the body and $\rho$ is its mean density. The dissipation function can also be defined in terms of the tidal Love number ($k_{\rm L}$) as $Q' = 3Q/2k_{\rm L}$. 

The tidal heating in Jupiter's moons Io and Europa works to circularize their orbits relative to Jupiter. Interestingly, estimates for their eccentricity damping timescales are considerably less than the age of the Solar System \citep[][p. 173]{murray1999}. Yet, their eccentricities are noticeably not zero (0.0043 and 0.0101, respectively). This inconsistency has been explained by the observed Laplace resonance between them and the satellite Ganymede \citep{peale1979,1981Icar...47....1Y}. The orbital periods of the three bodies are locked in a ratio of 1:2:4, so their gravitational interactions continually excite the orbits and maintain their non-zero eccentricities. This example demonstrates the need to include external gravitational influences when considering the long-term tidal evolution of non-isolated planet-moon systems. For systems in the HZ of low-mass star systems, external perturbations from the relatively close star could serve to maintain non-zero satellite eccentricity in much the same way as the orbital resonances in the Galilean moon system \citep{heller12}.

Equation~(\ref{eq:H}) represents the energy being tidally dissipated by the whole of a satellite. However, to assess the surface effects of tidal heating on a potential biosphere it is necessary to consider the heat flux through the satellite's surface. Assuming the energy eventually makes its way to the moon's surface, the tidal surface heat flux ($h$) can be represented as
\begin{equation} \label{eq:hconv} h_{\rm conv} = H/4{\pi}R_{\rm s}^2 \ . \end{equation}
Note in the definition that a subscript is used to identify Eq.~(\ref{eq:hconv}) as the surface heat flux based on the conventional tidal model for $H$. Below, we will define another tidal heat flux based on a different tidal model.

In terms of the effects of tidal heating on a moon's surface habitability, \cite{barnes09} proposed a conservative limit based on observations of Io's surface heat flux of about $h=2\,{\rm W\,m}^{-2}$ \citep{spencer00,2004jpsm.book..307M}, which results in intense global volcanism and a lithosphere recycling timescale on order of 10$^5$\,yr \citep{blaney1995,2004jpsm.book..307M}. Such rapid resurfacing could preclude the development of a biosphere, and thus $2\,{\rm W\,m}^{-2}$ can be considered as a pessimistic maximum rate of tidal heating to still allow habitable environments.

Tidal heating is not the only source of surface heat flow in a terrestrial body. Radiogenic heating, which comes from the radioactive decay of U, Th, and K, is an additional source for surface heat flow. \cite{barnes09} used this combination to also set a lower limit of $h_{\rm min} \equiv$ 0.04 W m$^{-2}$ for the total surface heat flux of a terrestrial body by considering that internal heating must be sufficient to drive the vitally important plate tectonics. This value was based on theoretical studies of Martian geophysics which suggest that tectonic activity ceased when the radiogenic\footnote{At the orbital distance of Mars, any contribution from tidal heating would be very low. Therefore, the total internal heating is essential equal to the radiogenic heating in this case.} heat flux dropped below this value \citep{williams1997}. Even though the processes which drive plate tectonics on Earth are not fully understood \citep{walker1981,regenauer01}, it is accepted that an adequate heat source is essential. The phenomenon of plate tectonic is considered important for habitability because it drives the carbon-silicate cycle thereby stabilizing atmospheric temperatures and CO$_2$ levels on timescales of $\sim$10$^8$\,yr. For reference, the Earth's combined outward heat flow (which includes both tidal and radiogenic heat) is 0.065 W m$^{-2}$ through the continents and 0.1 W m$^{-2}$ through the ocean crust \citep{zahnle07}, mostly driven by radiogenic heating in the Earth. 

Radiogenic heating scales as the ratio of volume to area \citep{barnes09}. Consequently, for most cases involving closely orbiting bodies that are significantly smaller than the Earth, it is believed that tidal heating probably dominates \citep{jackson08a,jackson08b}. The theoretical moons considered in this study meet these conditions. Therefore, we assume that radiogenic heating is negligible and that the total surface heat flux is equal to the tidal flux. We also adopt the heating flux limits for habitability presented above, so that $h_{\rm min} < h < h_{\rm max}$ could allow exomoon surface habitability.

\subsection{Global Energy Flux} \label{globflux}

Investigations of exomoon habitability can be distinguished from studies on exoplanet habitability in that a variety of astrophysical effects can be considered in addition to the illumination received by a parent star. For example, a moon's climate can be affected by the planet's stellar reflected light and its thermal emission.  Moons also experience eclipses of the star by the planet, and tidal heating can provide an additional energy source that is typically less substantial for planets. \cite{hb13} considered these effects individually, and then combined them to compute the orbit-averaged global flux ($F_{\rm glob}$) received by a satellite. More specifically, this computation summed the averaged stellar, reflected, thermal, and tidal heat flux for a satellite. In their study, they provided a convenient definition for the global flux as

\begin{equation}\label{Fglob}
F_{\rm glob} = \frac{L_\star(1-\alpha_s)}{16\pi a_{\rm \star p}^2\sqrt{1-e_{\rm \star p}^2}}\left( 1+\frac{\pi R_{\rm p}^2\alpha_{\rm p}}{2a_{\rm ps}^2}\right) + \frac{R_{\rm p}^2\sigma_{\rm SB}(T_p^{\rm eq})^4}{a_{\rm ps}^2}\frac{1-\alpha_s}{4} + h_s
\end{equation}

where $L_\star$ is the luminosity of the star, $a_{\rm \star p}$ is the semimajor axis of the planet about the star and $a_{\rm ps}$ is the satellite's semimajor axis about the planet, $\alpha$ is the Bond albedo, $e$ is eccentricity, $\sigma_{\rm SB}$ is the Stefan-Boltzmann constant, $h_s$ is the tidal heat flux in the satellite, and $T_{\rm p}^{\rm eq}$ is the planet's thermal equilibrium temperature.

As an analogy with the circumstellar HZ for planets, there is a minimum orbital separation between a planet and moon that will allow the satellite to be habitable. Moons inside this minimum distance are in danger of runaway greenhouse effects by stellar and planetary illumination and/or tidal heating. There is not a corresponding maximum separation distance (other than stability limits) because satellites with host planets in the stellar HZ are habitable by definition. The benefit of Eq.~(\ref{Fglob}) is that it can be used to explore the minimum distance. This is accomplished by comparing the global flux to estimates of the critical flux for a runaway greenhouse ($F_{\rm RG}$). \cite{hb13} discussed a useful definition for $F_{\rm RG}$ originally derived by \citet{2010ppc..book.....P}. Applying that definition to an Earth-mass exomoon gives a critical flux of $295\,{\rm W\,m}^{-2}$ for a water-rich world with an Earth-like atmosphere to enter a runaway greenhouse state. In comparison to the conservative limit of $2\,{\rm W\,m}^{-2}$ for the tidal heating discussed above, the runaway greenhouse limit defines the ultimate limit on habitability. Moons with a higher top-of-the-atmosphere energy flux cannot be habitable by definition, since all surface water will be vaporized.

\subsection{Coupling tidal and gravitational effects} \label{sim-evol}

\cite{heller12} suggested that low-mass stars cannot possibly host habitable moons in the stellar habitable zones because these moons must orbit their planets in close orbits to ensure Hill stability. In these close orbits they would be subject to devastating tidal heating which would trigger a runaway greenhouse effect and make any initially water-rich moon uninhabitable. This tidal heating was supposed to be excited, partly, by stellar perturbations. While tidal processes in the planet-moon system would work to circularize the satellite orbit, the stellar gravitational interaction would force the moon's orbital eccentricity around the planet to remain non-zero. However, \cite{heller12}  acknowledged that his model did not couple the tidal evolution with the gravitational scattering of a hypothetical satellite system so the extent of the gravitational influence of the star was surmised, but not tested. The need therefore remains to simulate the eccentricity evolution of satellites about low-mass stars with a model that considers both $N$-body gravitational acceleration and tidal interactions.

\textcolor{black}{We can estimate the stellar mass below which no habitable moons can exist for a given system age and planet mass. Neglecting all atmospheric effects on a water-rich, Earth-like object, we have
\begin{equation} \label{ahz}
a_{\rm HZ} \approx \left(\frac{L_\star}{L_\odot}\right)^{1/2} {\rm AU} \approx \left(\frac{M_\star}{M_\odot}\right)^{7/4} {\rm AU} \ ,
\end{equation}
\noindent
where $L_{\odot}$ and $M_{\odot}$ are the luminosity and mass of the Sun, respectively. The planetary Hill radius can be approximated as
\begin{equation}
R_{\rm Hill} \approx M_\star^{17/12} \left( \frac{M_{\rm p}}{3} \right)^{1/3} \frac{{\rm AU}}{M_\odot^{21/12}} \ .
\end{equation}
\noindent
Using the timescale for tidal orbital decay as used in Barnes \& O'Brien (2002; Eq. (7) therein), we have
\begin{equation} 
a_{\rm crit} = {\Bigg (} \frac{13 \tau_{\rm dec}}{2} \frac{3 k_{\rm 2,p} M_{\rm s} R_{\rm p}^5}{Q_{\rm p}} \sqrt{\frac{G}{M_{\rm p}}} + R_{\rm p}^{13/2} {\Bigg )}^{2/13} \ \stackrel{\displaystyle !}{<} \ R_{\rm Hill} \ ,
\end{equation}
\noindent
for tidal love number $k_{\rm 2,p}$ for the planet, where the latter relation must be met to allow for tidal survival over a time $\tau_{\rm dec}$. Hence
\begin{equation} \label{minstarmass}
M_\star \stackrel{\displaystyle !}{>} {\Bigg (} a_{\rm crit} \left( \frac{3}{M_{\rm p}} \right)^{1/3} \frac{M_\odot^{7/4}}{\rm AU} {\Bigg )}^{12/17} \ .
\end{equation}
\noindent
For a Jupiter-like planet ($Q_{\rm p}=10^5$, $k_{\rm 2,p}=0.3$) with an Earth-like moon, and assuming tidal survival for 4.5\,Gyr, we estimate a minimum stellar mass of $0.18\,M_\odot$.}

For a computational study, many popular and well tested computer codes are available to simulate the gravitational (dynamical) evolution of many bodied systems \citep{chambers1997,rauch02}. Such codes are particularly useful for studying the long-term stability of planetary systems. These codes, however, do not include tidal interactions in their calculations. A modification of the \textit{Mercury} $N$-body code \citep{chambers1997} to include tidal effects \citep{2015A&A...583A.116B} has recently been used to simulate, for the first time, the evolution of multiple moons around giant exoplanets \citep{heller14}. Yet, these orbital calculations neglected stellar effects in the planet-moons system.

Useful derivations for a tide model that provides the accelerations from tidal interactions at any point in a satellites orbit derivations were presented by \citet{eggleton1998}, whose work was based on the equilibrium tide model by \citet{hut1981}. Their particular interest was to consider tidal interactions between binary stars. \cite{eggleton1998} derived from first principles equations governing the quadrupole tensor of a star distorted by both rotation and the presence of a companion in a possibly eccentric orbit. The quadrupole distortion produces a non-dissipative acceleration $\bm{f}_{\rm QD}$. They also found a functional form for the dissipative force of tidal friction which can then be expressed as the acceleration due to tidal fiction $\bm{f}_{\rm TF}$. These acceleration terms are useful because they can be added directly to the orbital equation of motion for the binary:

\begin{equation} \label{eofm}
\ddot{\bm{r}} = -\frac{GM\bm{r}}{r^3} + \bm{f}_{\rm QD} +  \bm{f}_{\rm TF}\ \ ,
\end{equation}

where $\bm{r}$ is the distance vector between the two bodies and $M$ is the combined mass. This enables the evaluation of both the dynamical and tidal evolution of a binary star system.

\subsection{A Self-Consistent Evolution Model\\* for Planetary Systems} \label{method}

\cite{mardling02} were the first to recognize that the formulations by \cite{eggleton1998} provided a powerful method for calculating the complex evolution of not just binary stars, but planetary systems as well. Based on their formulations,  \cite{mardling02} presented an efficient method for calculating self-consistently the tidal plus $N$-body evolution of a many-bodied system. Their work had a particular focus on planets, yet they emphasized that the method did not assume any specific mass ratio and that their schemes were entirely general. As such, they could be applied to any system of bodies.

The \citet{mardling02} method lends itself best to a hierarchical (Jacobi) coordinate system. Since our primary interest involves the evolution of moons around a large planet, we use the planet's position as the origin of a given system. The orbit of the next closest body will be a moon, whose position is referred to the planet. The orbit of a third body is then referred to the center of mass of the planet and innermost moon, while the orbit of a fourth body is referred to the center of mass of the other three bodies. This system has the advantage that the relative orbits are simply perturbed Keplerian orbits so the osculating orbital elements are easy to calculate \citep{murray1999}.

\cite{mardling02} parameterized the acceleration terms to create equations of motion for systems containing up to four bodies. Let the masses of the four objects be $m_1, m_2, m_3$, and $m_4$. For our study, $m_1$ always represented a planet and $m_2$ represented a moon. The third mass, $m_3$, represented a central star for 3-body systems, which are created by simply setting $m_4$ equal to zero. The planet and moon (being the closest pair of bodies in the system) are endowed with structure that is specified by their radii $S_1$ and $S_2$, moments of inertia $I_1$ and $I_2$, spin vectors ${\bm{\Omega}}_1$ and ${\bm{\Omega}}_2$, their quadrupole apsidal motion constants (or half the appropriate Love numbers for planets with some rigidity) $k_1$ and $k_2$, and their $Q$-values $Q_1$ and $Q_2$. Body 3 is assumed to be structureless, meaning it is treated as a point mass. \textcolor{black}{We should note one necessary correction to Eq.~(7) in \cite{mardling02}, the $m_3\bm{\beta}_{34}$ term in the first set of brackets should be replaced with $m_4\bm{\beta}_{34}$.}

The total angular momentum and total energy for each system were calculated as well as the evolution of the spin vectors for bodies 1 and 2 \citep[Eqs.~9, 10, and 15 in][]{mardling02}. We should also note one minor correction to Eq.~(14) in \cite{mardling02}, the factor $S_1^5$ should be replaced by $S_2^5$.
%

The acceleration from quadrupole distortion $\bm{f}_{\rm QD}$ was derived from a potential and under prevailing circumstances would conserve total energy. \textcolor{black}{On the other hand, the acceleration due to tidal friction $\bm{f}_{\rm TF}$ cannot be derived from a potential and represents the effects of a slow dissipation of orbital energy. The dissipation within the moon causes change in its orbital semi-major axis and spin vector. We defined the rate of energy loss from tidal heating $\dot{E}_{\rm tide}$ using Eq.~(71) in \cite{mardling02}. While total orbital energy is not conserved, we still expect conservation of total angular momentum.} Since $\dot{E}_{\rm tide}$ represents the tidal heating in a satellite according to this tidal model, the surface heat flux in the satellite can be defined as 
\begin{equation} \label{surfheat} h = \dot{E}_{\rm tide}/4{\pi}S_{2}^2 \ . \end{equation}

\subsubsection{The Simulation Code}

Using the equations of motion defined in the previous subsection, we \textcolor{black}{designed} a computer program that has the ability to simultaneously consider both dynamical and tidal effects, and with it, we simulated exomoon tidal evolution in low-mass star systems. The program code was written in C++ and a Bulirsch-Stoer integrator with an adaptive timestep \citep{nrcplus} was used to integrate the equations of motion.

 The robustness of the code was evaluated by tracking the relative error in total angular momentum, defined as $(L_{\rm out} - L_{\rm in})/L_{\rm in}$, where $L_{\rm in}$ is the system angular momentum at the start of a simulation and $L_{\rm out}$ is the angular momentum at a later point. The size of the relative error could be controlled by adjusting an absolute tolerance parameter. However, as is often the case with direct orbit integrators, the problem of systematic errors in the semi-major axis existed. \textcolor{black}{The simulations were monitored to ensure a maximum allowed error of 10$^{-9}$ for the total angular momentum.} The algorithm efficiency required about seven integration steps per orbit, for the smallest orbit. Relative errors in total energy ($(E_{\rm out} - E_{\rm in})/E_{\rm in}$) were also tracked, although, conservation of mechanical energy was not expected for our systems and, hence, not included as a performance constraint in our simulations.

A primary drawback to directly integrating orbital motion is the extensive computational processing times required to simulate long-term behavior.  With our particular evolution model the situation is compounded by extra calculations for tidal interactions. Early tests for code efficiency indicated a processing time of roughly two days per Myr of simulated time for 3-body systems (two extended objects and a third point mass object). \textcolor{black}{Integration time significantly increases when additional bodies are considered, so a maximum of 3 bodies was chosen for this initial study.} Since the integration timestep is effectively controlled by the object with the shortest orbital period, the exact processing times varied significantly between wide orbit and short orbit satellites. For our study the size of the orbit was determined from the theoretical habitable zone around a low-mass star (see subsection \ref{planet_model}). From these results we predicted an ability to simulate satellite systems with timescales on order of 10$^7$\,yr with our limited computational resources.

\section{Simulating Exomoon Evolution: Setup}\label{chap5}

\subsection{System Architectures and Physical Properties} \label{archit}

As part of our investigation into the evolution of exomoons around giant planets in the HZ of low-mass stars, we evaluated two different system architectures. The first involved a minimal 2-body system consisting of only a planet and a moon; the second was a 3-body system of one planet, one moon, and a central star. For each system, the planet and moon were given structure while a star was treated as a point mass (which significantly reduce the required number of calculations per timestep). For each 3-body simulation, a corresponding 2-body simulation was performed for the same planet-moon binary.  Comparing the two simulations would provide a baseline for determining the stellar contribution to a moon's long-term evolution. 

Although thousands of extrasolar planets have been detected, very little information is known about their internal structure and composition. \textcolor{black}{Notwithstanding, work is being conducted to offer a better understanding (for example, see \cite{unterborn16} and \cite{dorn17}).} Recognizing that limits for tidal dissipation depend critically on these properties, without this information, objects in the solar system provide the best guide for hypothesizing the internal structure and dynamics of extrasolar bodies. For this reason, we used known examples from our Solar System to model the hypothetical extrasolar satellite systems.

\subsubsection{The Exomoon Model} \label{moon_model}

Formation models for massive exomoons show reasonable support for the formation of moons with roughly the mass of Mars around super-Jovian planets \citep{canup06,2015ApJ...806..181H}, which is near the current detection limit of $\sim$0.1 $M_{\oplus}$ \citep{2014ApJ...787...14H,2015ApJ...813...14K} and also lies in the preferred mass regime for habitable exomoons (see Sect.~\ref{intro}). For these reasons, we chose to model the physical structure of our hypothetical exomoons after planet Mars. 

\begin{table}
  \caption{Physical properties for a hypothetical Mars-like exomoon. The parameters $A, B$ and $C$ are the principal moments of inertia. \vspace{3mm}}
  \label{tab:mars}
  \begin{center}
    \begin{tabular}{|l|c|}
      \hline 
      \textbf{Parameter} & \textbf{Value} \\\hline
      Mass ($M$) & 0.107 $M_{\oplus}$ \\
      Mean Radius ($R$) & 0.532 $R_{\oplus}$\\
      Love Number ($k_{\rm L}$) & 0.16\\
      Bond Albedo ($\alpha$) & 0.250\\
      Dissipation Factor ($Q$) & 80\\
      $C/MR^2$ & 0.3662\\
      $C/A$ & 1.005741\\
      $C/B$ & 1.005044\\
      \hline
    \end{tabular}
  \end{center}
\end{table}

The specific physical properties used in our moon model are shown in Table \ref{tab:mars}. The Bond albedo represents current estimates for Mars, although one could argue that an Earth-like value of 0.3 might be equally appropriate since we consider habitable exomoons. Bond albedos are not actually used in the evolution simulations, they are utilized afterwards to estimate the global flux $F_{\rm glob}$ received by the moon, see Eq.~(\ref{Fglob}). Habitability considerations from the global flux are made in comparison to the critical flux for a runaway greenhouse ($F_{\rm RG}$, see subsection \ref{globflux}). For a Mars-mass exomoon, the critical flux is $F_{\rm RG} = 269\,{\rm W\,m}^{-2}$. We decided to use the lower Bond albedo of Mars, keeping in mind that it would produce slightly higher estimates of $F_{\rm glob}$. If the simulation results showed a global flux $>269\,{\rm W\,m}^{-2}$ we would then also consider an Earth-like Bond albedo.

The Love number and dissipation factor are also based on recent estimates for Mars \citep{yoder03,bills05,lainey07,konopliv11,nimmo13}. Their specific values represent the higher dissipation range of the estimates. This choice produces a slightly faster tidal evolution and also tests the extent of the gravitational influence of the star against a slightly higher rate of energy loss that continually works to circularize the orbit of the moon. The principal moments of inertia also reflect estimates for Mars \citep{bouquillon1999} and represent a tri-axial ellipsoid for the overall shape of the body. We realize that a Mars-like exomoon orbiting close to a giant planet would undoubtedly develop a different bodily shape than the current shape of Mars. If we assume a constant moment of inertia, then the principal moments only become important in calculating the evolution of the moon's spin vector \citep[see Eq. 15 in][]{mardling02}. To minimize this importance, we set the moon's initial obliquity to zero and started each simulation with the moon in synchronous rotation. Under these conditions there is minimal change to the moon's spin vector and the given triaxial shape is effective at keeping the moon tidally locked to the planet as its orbit slowly evolves. In that sense, another choice in shape could have equally served the same purpose. 

We only considered prograde motion for the moon relative to the spin of the planet. In keeping with our use of local examples, we modeled the moon's orbital distance after known satellite orbits around giant planets in the Solar System. The large solar system moons Io, Europa, Ganymede, Titan, and Callisto happen to posses roughly evenly spaced intervals for orbital distance in terms of their host planet's radius $R_{\rm p}$ (5.9, 9.6, 15.3, 21.0, and 26.9 $R_{\rm p}$, respectively). In reference to this natural spacing, when discussing orbital distances of a moon we will often refer to them as Io-like or Europa-like orbits, etc. 

Considering stability constraints, and assuming our satellite systems are not newly formed, we would not expect high eccentricities for stable exomoon orbits. For this reason we use an initial eccentricity of 0.1 for our 2-body and 3-body models. We then monitor the moon's evolution as the tidal dissipation works to circularize the orbit.

\subsubsection{The Planet Models}\label{planet_model}
Moon formation theories suggest that massive terrestrial moons will most likely be found around giant planets. With this consideration, we chose to model our hypothetical planet after the two most massive planets in our Solar System, specifically, planets Jupiter and Saturn. The physical properties used in our planet models are shown in Table \ref{tab:pl_props}. The Bond albedos, Love numbers, and dissipation factors are based on current estimates \citep{gavrilov1977,hanel1981,hanel1983,meyer07,lainey09}. We chose values that represent more substantial heating in the planets, similar to our choice for the Mars-like moons. 

The normalized moments of inertia were also based on recent estimates \citep{helled11sat,helled11jup} and the three equal principal moments imply a spherical shape for the planet. This shape may be unrealistic in that a planet orbiting close to a star with a massive moon is unlikely to maintain a truly spherical shape. With spherical bodies we also do not match the exact shapes of the solar system planets, i.e., we ignore their oblateness that results from their short rotation periods (about 10 hrs for Jupiter, 11 hrs for Saturn). However, the planet's exact shape is not particularly import for the purposes of our simulations. We start each planet with zero obliquity and synchronous rotation relative to its orbit around the central star, which results in a rotation period that ranges from about 15 to 120 days. This setup is consistent with the prediction that planets in the HZ of low-mass stars will most likely be tidally locked to the star.

\begin{table}  
  \caption{Physical properties for hypothetical giant exoplanets. The planet shape is assumed spherical with principal moments of inertia $A = B = C$. \vspace{3 mm}}
  \label{tab:pl_props}
  \begin{center}
    \begin{tabular}{|l|c|}
      \hline 
      \multicolumn{2}{|c|}{\textbf{Jupiter-like}}\\ \hline
      \textbf{Parameter} & \textbf{Value} \\\hline
      Mass ($M$) & 318 $M_{\oplus}$ \\
      Mean Radius ($R$) & 11.0 $R_{\oplus}$\\
      Bond Albedo ($\alpha$) & 0.343\\
      Love Number ($k_{\rm L}$) & 0.38\\
      Dissipation Factor ($Q$) & 35000\\
      $C/MR^2$ & 0.263\\
      \hline
    \end{tabular}
    \quad
    \begin{tabular}{|l|c|l|}
      \hline 
      \multicolumn{2}{|c|}{\textbf{Saturn-like}} \\ \hline
      \textbf{Parameter} & \textbf{Value} \\\hline
      Mass ($M$) & 95.2 $M_{\oplus}$ \\
      Mean Radius ($R$) & 9.14 $R_{\oplus}$\\
      Bond Albedo ($\alpha$) & 0.342\\
      Love Number ($k_{\rm L}$) & 0.341\\
      Dissipation Factor ($Q$) & 18000\\
      $C/MR^2$ & 0.21\\
      \hline
    \end{tabular}
  \end{center}
\end{table}

To define the HZ for a given star we followed the work by \cite{kopparapu13a}, who estimated a variety of limits for the inner and outer edges around stars with effective temperatures ($T_{\rm eff}$) in the range 2600 K $\leq~T_{\rm eff}~\leq$ 7200 K. For this study we used their conservative limits corresponding to a moist greenhouse for the inner boundary and a maximum greenhouse for the outer boundary. In order to calculate these limits it is necessary to define the effective temperature and radius of the star. For our purposes, however, it was more convenient to scale the HZ based on stellar mass. The stellar radius can be estimated from the mass using an empirical relation derived from observations of eclipsing binaries \citep{gorda1999}:
\begin{equation} \label{eq:radrel}  {\log}_{10}\frac{R_\star}{R_{\odot}} = 1.03\ {\log}_{10}\frac{M_\star}{M_{\odot}} + 0.1. \end{equation}
Note that Eq.~(\ref{eq:radrel}) is valid only when $M_\star \lesssim M_{\odot}$, which is valid for the dwarf stars we consider. The stellar luminosity can be estimated from the mass by

\begin{equation} \lambda = 4.101{\mu}^3 + 8.162{\mu}^2 + 7.108{\mu} + 0.065, \end{equation}

where $\lambda = \log_{10}(L/L_{\odot})$ and $\mu = \log_{10}(M_\star/M_{\odot})$ \citep{scalo07}. With luminosity and radius, the stellar effective temperature can be calculated using the familiar relationship

\begin{equation} \label{eq:starl} L = 4{\pi}{\sigma_{\rm SB}}R_\star^2\ T_{\rm eff}^4 . \end{equation}

\textcolor{black}{An online database of confirmed exoplanet detections\footnote{NASA Exoplanet Archive at \href{https://exoplanetarchive.ipac.caltech.edu/}{exoplanetarchive.ipac.caltech.edu/}} lists almost 3,500 confirmed planets in total. Of those, only about 8 percent have host stars with mass $\lesssim 0.6\ M_{\odot}$. This relatively small sample is most likely due to selection bias as detection techniques have evolved. The majority of these detections occurred in recent years and the number is expected to grow. While several large planets can be found orbiting low-mass stars, only a small number orbit inside the HZ. The percentage of these systems is questionable since a significant number of confirmed planets in the database are missing key parameters such as planet mass or orbit distance.} One known planet in particular corresponds nicely with our Saturn model. The planet HIP 57050 b has a mass of 0.995 $(\pm0.083)\ M_{\rm Sat} \times \sin(i)$ \citep{2010ApJ...715..271H}, where $M_{\rm Sat}$ is the mass of Saturn and $i$ is the unknown inclination between the normal of the planetary orbital plane and our line of sight. Since no other information is available about the physical properties of these giant planets, only direct comparisons relating to their masses can be made. None of the \textcolor{black}{low-mass} HZ candidates matches directly with Jupiter, but two have masses about double that of Jupiter (GJ 876 b and HIP 79431 b). This at least supports the plausibility for the existence of Jupiter mass planets in the HZ of dwarf stars.

Each simulated system consisted of only one planet. Each planet started with a circular orbit relative to the star at a distance inside the stellar HZ. To conserve computational processing time, we limited exploration of the HZ to just two specific orbital distances per planetary system. The first location was in the center of the HZ for a given star mass, otherwise defined as 

\begin{equation} \label{cendis} a_{\rm center} = (\textrm{inner edge} + \textrm{outer edge})/2 \ .\end{equation}

This particular location was to serve as a reference point for the next round of simulations. If most of our hypothetical satellites already experienced intense tidal heating at the center, then the next location should be further out in the zone. On the other hand, if the surface heating rates were below the proposed maximum for habitability ($h_{\rm max}=2\,{\rm W\,m}^{-2}$), then we would move inward for the next round. As we will show, after simulating satellite systems in the center of the HZ it became clear that the second round of simulations should involve the inner HZ.

While the innermost edge was a reasonable option to explore, an Earth-equivalent distance had obvious attraction. By `Earth-equivalent' we refer to the Earth's relative position in the Sun's HZ as compared to the total width of the zone. Conservative estimates by \cite{kopparapu13a} place the inner edge of the Sun's HZ at 0.99\,AU and the outer edge at 1.67\,AU. With these boundaries we define Earth's relative location within the solar HZ as

\begin{equation} d_{\rm rel} = 1 - (1.67\,{\rm AU} - 1\,{\rm AU})/(1.67\,{\rm AU} - 0.99\,{\rm AU}) = 0.0147 \ .\end{equation}

For a given star, we use this relative location and the width of its HZ to define a planet's Earth-equivalent orbital distance as

\begin{equation} \label{eqdis} a_{\rm eq} = \textrm{inner edge} + (\textrm{outer edge} - \textrm{inner edge}) \ \times \ d_{\rm rel} \ .\end{equation}

\subsubsection{The Star Models}
Information relating to a star's mass, radius, and effective temperature is required to calculate its circumstellar HZ and to estimate the global flux received by a moon. In our simulations the stars are treated as point masses, and so the stellar mass is the only physical characteristic used in the actual simulation of a system. For our study we considered a star mass range from 0.075 $M_{\odot}$ to 0.6 $M_{\odot}$. \textcolor{black}{These would consist of mostly M spectral type stars and possibly some late K type stars, with typical surface temperatures less than 4,000 K. This range of stars is sometimes referred to as red dwarf stars.}

\subsection{Estimates for Eccentricity Damping Timescales} \label{escales}

We can estimate the eccentricity damping timescales ($\tau_e$) for our hypothetical satellite systems following Eq.~(\ref{tdamp}). The timescales are summarized in Table \ref{tab:dampscales} for the two planet models and for the various moon orbital distances considered. It was fortunate to see estimates around 1\,Myr since integration times of order $10^7$\,yr represent achievable computational processing times. Results showed that we were able to simulate complete tidal evolution for moons with Io-like and Europa-like orbital distances. However, it was unlikely that we could show any significant evolution for the wider orbits on these timescales. It should be noted that Eq.~(\ref{tdamp}) was derived from a two body calculation. As such, it does not take into account any perturbing effects from additional bodies. Since the degree to which the central star would influence a moon's orbit was unknown, we decided to include the wider orbits in our considerations.

\begin{table} 
  \caption{Eccentricity damping timescale estimates for a Mars-like moon at different orbital distances. \vspace{3 mm}}
  \label{tab:dampscales}
  \begin{center}
    \begin{tabular}{|l|c|}
      \hline 
      \multicolumn{2}{|c|}{\textbf{Jupiter-like Host Planet}}\\ \hline
      \textbf{Orbital Distance} & $\boldsymbol{\tau_e}$ \textbf{ (years)} \\\hline
      Io-like ($5.9 R_{\rm Jup}$)  &  4 x $10^5$\\
      Europa-like ($9.6 R_{\rm Jup}$) & 9 x $10^6$\\
      Ganymede-like ($15.3 R_{\rm Jup}$)& 2 x $10^8$\\
      Titan-like ($21.0 R_{\rm Jup}$)& 1 x $10^9$\\
      Callisto-like ($26.9 R_{\rm Jup}$)& 7 x $10^9$\\
      \hline
    \end{tabular}
    \quad
    \begin{tabular}{|l|c|l|}
      \hline 
      \multicolumn{2}{|c|}{\textbf{Saturn-like Host Planet}}\\ \hline
      \textbf{Orbital Distance} & $\boldsymbol{\tau_e}$ \textbf{ (years)} \\\hline
      Io-like ($5.9 R_{\rm Sat}$) &  6 x $10^5$\\
      Europa-like ($9.6 R_{\rm Sat}$) & 1 x $10^7$\\
      Ganymede-like ($15.3 R_{\rm Sat}$)& 3 x $10^8$\\
      Titan-like ($21.0 R_{\rm Sat}$)& 2 x $10^9$\\
      Callisto-like ($26.9 R_{\rm Sat}$)& 1 x $10^{10}$\\
      \hline
    \end{tabular}
  \end{center}
\end{table}

\section{2-Body Orbital Evolution: Tides versus No-Tides} \label{tvsnot}

\begin{figure}
     \begin{center}
        \subfigure[]{
            \label{fig:2Be}
              \includegraphics[width=.48\textwidth]{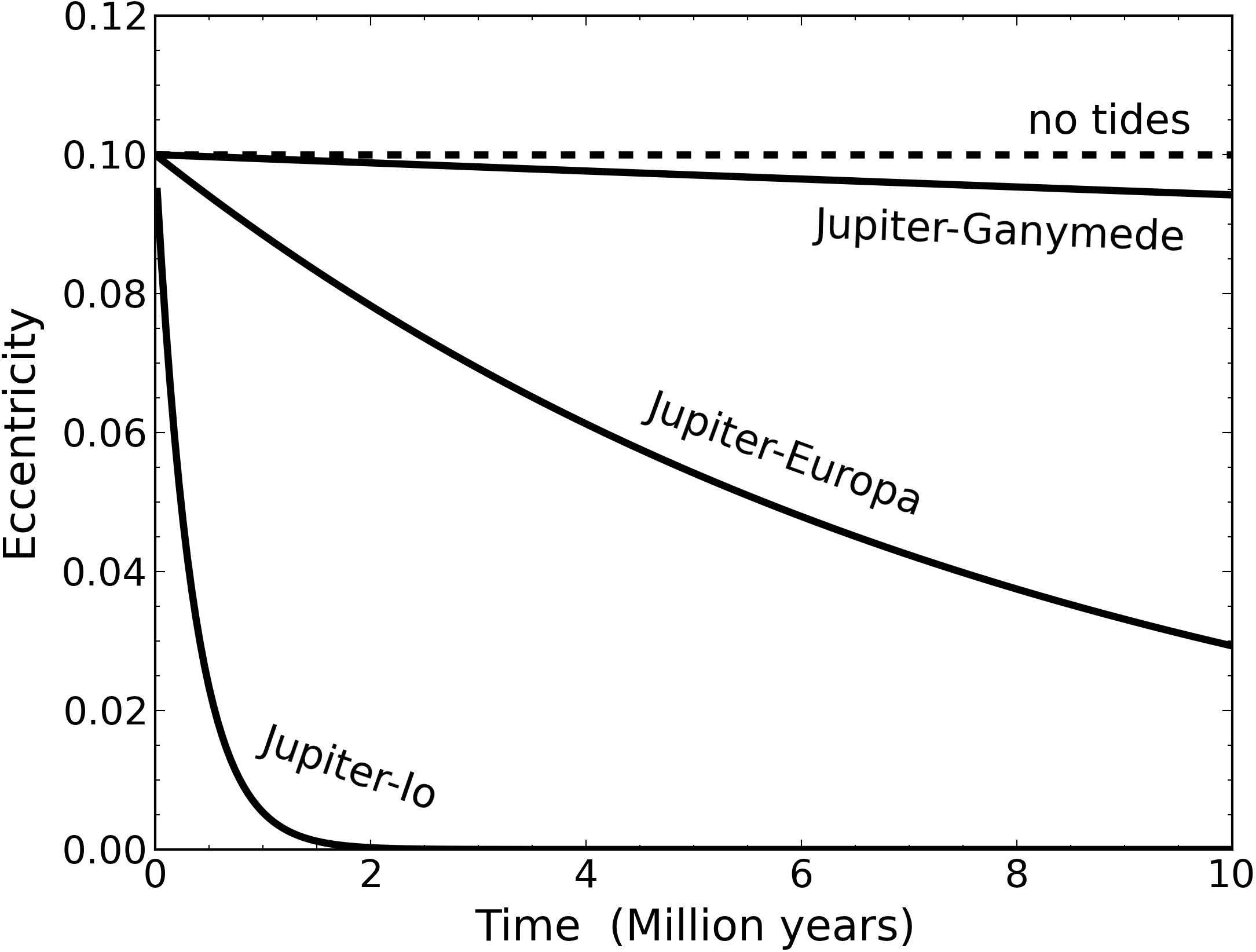}
        }
       \subfigure[]{
            \label{fig:2Ba}
              \includegraphics[width=0.48\textwidth]{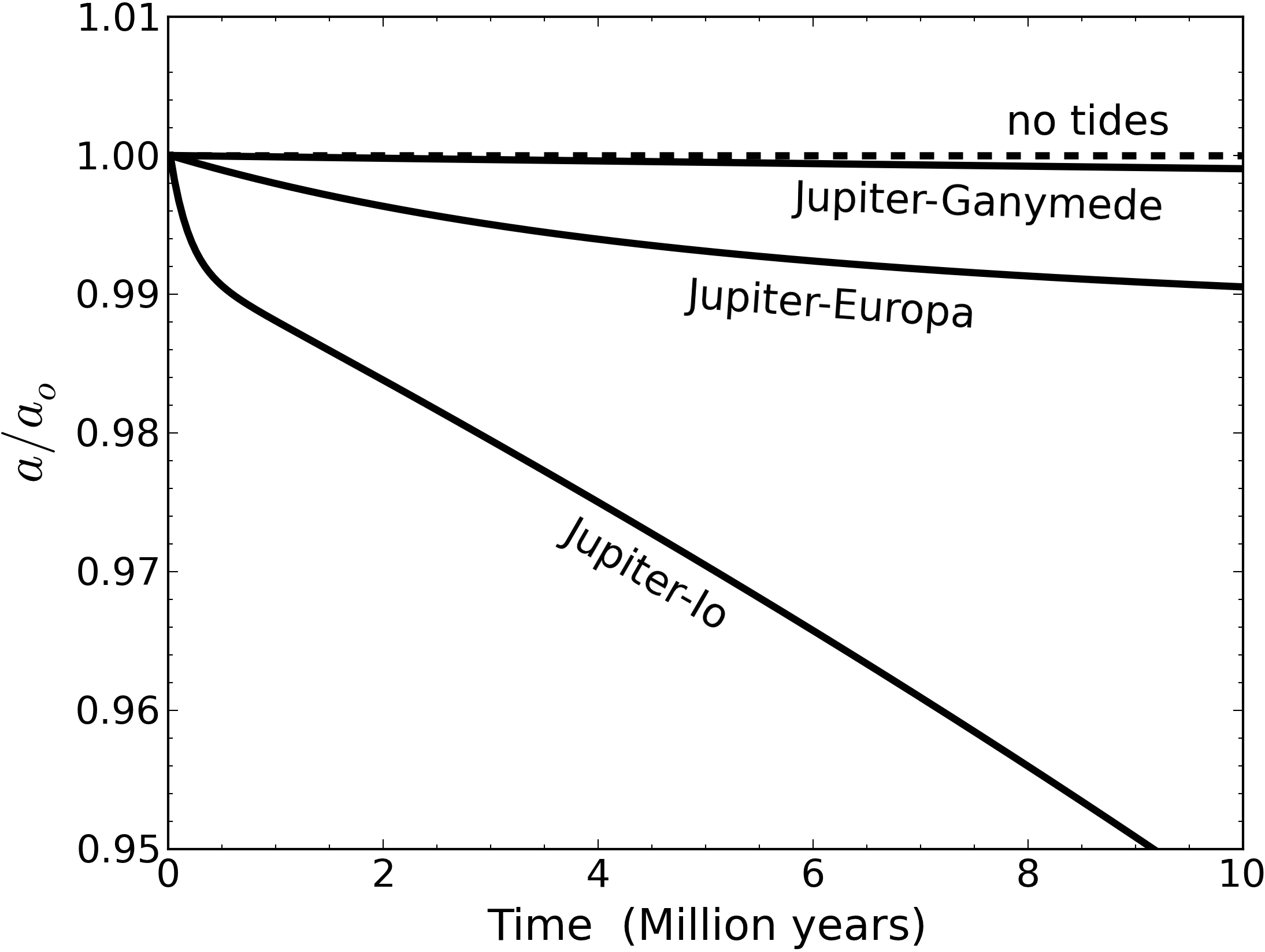}
        }
    \end{center}
    \caption{Orbital evolution of a single Mars-like moon around a Jupiter-like planet in four different cases. The dashed line assumes no tides. The solid lines all assume an initial eccentricity of 0.1 but with the moon starting at different semi-major axes from the planet, i.e. in an Io-wide, a Europa-wide, and in a Ganymede-wide orbit around the planet (see labels).}
   \label{fig:shortjup2b}
\end{figure}

Figure~\ref{fig:shortjup2b} demonstrates some typical evolutions for Mars-like moons around a Jupiter-like planet. Solid curves are simulations that included tidal interactions. The lower, middle, and upper red curves represent Io-like, Europa-like, and Ganymede-like moon orbital distances, respectively. As expected, the rate of change depended strongly on the moon's orbital distance. When repeated without tidal interactions, all simulations had the same result, with no significant change to the orbits (see dashed lines). Simulations for Titan-like and Callisto-like orbital distances showed very little change over 10\,Myr whether tides were considered or not. Their results are nearly identical to the dashed curves in each plot so they were not included in Fig.~\ref{fig:shortjup2b}.

Simulations for a Saturn-like host planet were also conducted. The satellite evolutions are similar to those in Fig.~\ref{fig:shortjup2b} for a Jupiter-like host planet, with one exception. An Io-like moon orbit around Saturn experienced significantly greater change in semi-major axis than the Io-like moon orbit around Jupiter. In just 7\,Myr the semi-major axis decayed to 75\% of its initial value with a Saturn-like host planet, compared to about 96\% with a Jupiter-like host. The explanation is related to our approach of normalizing planet-moon distances by the radius of the respective host planet. A moon's initial orbital distance is defined by $R_{\rm p}$, the planet radius (an `Io-like' orbit is 5.9 $R_{\rm p}$). Because of Saturn's smaller radius a moon with an Io-like orbital distance is actually closer to the planet than an Io-like orbit around Jupiter, since 5.9 $R_{\rm Sat} < $ 5.9 $R_{\rm Jup}$. Following this formulation for the orbital distances, identical moons around each planet will experience different heating rates.

While different tidal heating rates explain part of the discrepancy in semi-major axis evolution between Saturn-like and Jupiter-like host planets, it does not account for all of it. Figure~\ref{fig:2Be} shows the Io-like orbit around the Jupiter-like host was circularized after $\approx2.5$\,Myr, at which point the tidal heating ceases in our synchronized moon with zero spin-orbit alignment. Therefore, any change in the semi-major axis after $\approx2.5$\,Myr is no longer attributed to tidal dissipation in the moon. Instead, tidal dissipation must occur in the planet, which itself is synchronized to the star, causing the planet's tidal bulge to lag behind the moon in our systems. This continued evolution makes the moon spiral towards the planet. Ultimately, the moon will be destroyed near the planetary Roche radius \citep{barnes02}. Tidal evolution of moons in Europa-wide orbits are much slower, with a fractional decrease of the semi-major axis of $\approx0.1\,\%$ per 10\,Myr. 

Using the corresponding slope in Fig.~\ref{fig:2Ba} for the Jupiter system, we roughly estimate a total lifetime of $\approx200$\,Myr for the moon with an Io-like orbit. \textcolor{black}{As a comparison of this timescale to theoretical predictions, \cite{gs66} derived a relation for the change in semi-major axis of a satellite as
\begin{equation} \label{deltaa} \frac{da}{dt} = \frac{9}{2} \left(\frac{G}{M_p}\right)^{1/2} \frac{R_p^5}{Q'_p} \frac{M_s}{a^{11/2}} \ .\end{equation}
This model assumes that the planet's $Q$-value is frequency-independent, which is only a fair approximation over a very narrow range of frequencies and therefore implicitly limits the model to low eccentricities and inclinations \citep{2009ApJ...698L..42G}.} 

\textcolor{black}{As an alternative, \cite{Mardling2011} suggested to use the constant time lag ($\tau$) model to study the tidal evolution of hot Jupiters based on the knowledge of tidal dissipation in Jupiter. This is done by assuming that $\tau$ (rather than $Q$) is common to Jupiter-mass planets. \citet{Mardling2011} propose to approximate the $Q$-value of a planet with orbital period $P$ via $Q/Q_{\rm J} = P/P_{\rm Io}$, where $Q_{\rm J}$ and $P_{\rm Io}=1.77$\,d are the tidal dissipation constant of Jupiter and the orbital period of Io, respectively. The shortest orbital period for our Jupiter-like planets was 13\,d, which gives us an estimated $Q$-value 7.3 times that for Jupiter. Applying this result  and using the values listed in Table \ref{tab:pl_props}, Eq.~(\ref{deltaa}) estimates a timescale of $\approx560$\,Myr for the orbital decay. While a few times longer than the rough estimate from our simulations, they agree on the order of magnitude.}



\begin{figure}
     \begin{center}
        \subfigure[]{
            \label{fig:spincomp}
            \includegraphics[width=.5\textwidth]{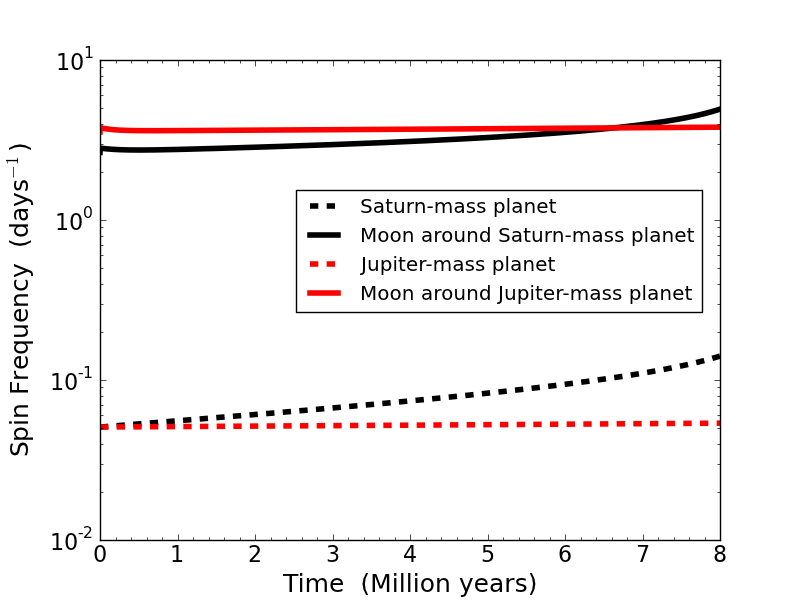}
        }
        \subfigure[]{
          \label{fig:spinmean}
            \includegraphics[width=.5\textwidth]{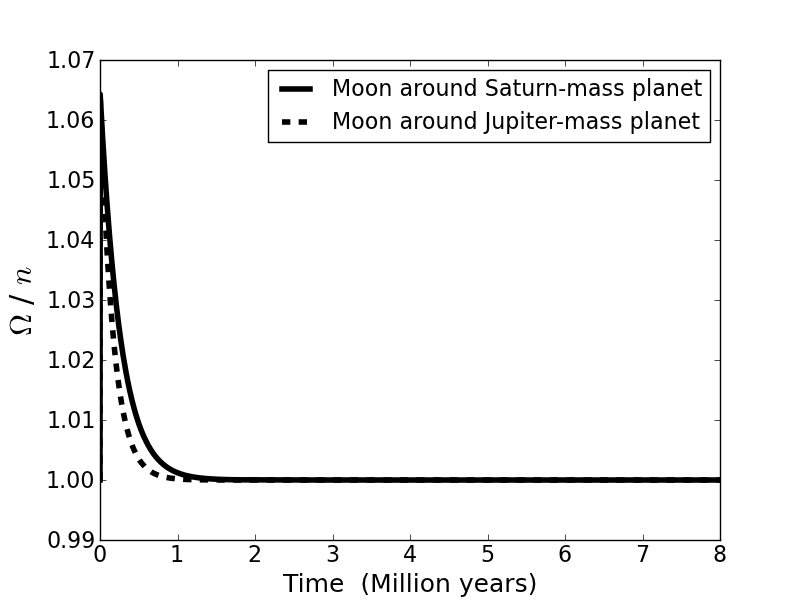}
        }
    \end{center}
    \caption{Spin evolution of a moon and its host planet, assuming an Io-like moon orbit. (\textbf{a}) Spin frequencies of the planets (dashed lines) and moons (solid lines) assuming host planets akin to Jupiter (red) and Saturn (black). (\textbf{b}) Ratio of spin magnitude ($\Omega$) and mean motion ($n$) for the moons represented in plot a.}
   \label{fig:spinevo}
\end{figure}

A graphical representation of the spin evolution for the Io-like systems is provided in Fig.~\ref{fig:spinevo}. Comparing the planet spin rates (dashed lines) in Fig.~\ref{fig:spincomp}, the Saturn-like planet (black dashed line) had a noticeable increase relative the Jupiter-like host (red dashed line), note the log scaling for the ordinate axis. This difference is explained by the unequal distances for the otherwise identical Mars-like moons as well as the unequal moments of inertia for the planets. The short satellite orbit around Saturn combined with its lower moment of inertia leads to the small, but noticeable increase in spin. That small increase causes the moon to almost double its initial spin rate. On the other hand, the wider orbit and more massive Jupiter host planet caused very little change for both the moon and planet in this system. 

The tidally locked state of the moons throughout the simulated time are demonstrated in Fig.~\ref{fig:spinmean}. A perfectly synchronized spin rate would produce a value of 1 for the ratio between spin magnitude and mean orbital motion. Notice that the ratio is not exactly 1 for the first few Myr, although the difference is small. This discrepancy can be explained by the shapes of the orbit during that time. Referring to the eccentricity plotted in Fig.~\ref{fig:2Be}, the spin cannot completely synchronize with the orbital period until the eccentricity approaches zero.

As a final observation, one similarity between the Jupiter and Saturn-like systems is that there is little change in the orbital elements over a period of 10\,Myr for Ganymede-like orbits and greater. This was not particularly surprising as the result was predicted from the damping timescales. At these wider orbits the dissipation rate would be much lower for a given eccentricity due to the $a^{-15/2}$ dependence of tidal heating. On the other hand, a central star's potential for exciting the eccentricity is still untested, so we included the wider orbits in our 3-body analysis.

\section{3-Body Simulations} \label{chap6}

\textcolor{black}{We continue with systems consisting of one star, one planet, and one moon. The Mars-like moon was given an initial eccentricity of 0.1, measured relative to the planet. We used this high value for eccentricity to cover a wide range of formation possibilities and allowed the eccentricity to decay as a result of tidal dissipation in the moon and planet. We monitored the simulations to see if the moon orbits would settle to steady, non-zero eccentricities long after the orbits should have circularized due to the tidal heating.} For each satellite system we also ran a nearly identical simulation with the moon instead starting with a circular orbit. In this way we could test if the stellar perturbation raised the eccentricity to the same steady state value as was achieved following the decay from a higher initial value. For each 3-body simulation, we ran a corresponding 2-body simulation with just the planet and moon to show what the evolution would be without the influence of the star.

\subsection{3-body Stability Considerations}

\begin{table}
  \caption{3-body stability summary for a Jupiter-like host planet. The moon semi-major axes are presented as fractions of $R_{\rm Hill}$. Gray shaded cells represent unstable moon orbits.}
  \label{tab:stab3brhilljup}
  \begin{center}
    \vspace{2 mm}
    \begin{tabular}{|c|c|c|c|c|c|}
      \hline 
      \multicolumn{6}{|c|}{\textbf{Earth-Equivalent Planet Orbit}}\\ \hline 
      \multirow{2}{*}{Star Mass} & \multicolumn{5}{|c|}{Moon Semi-major Axis}\\ \cline{2-6}
      & {5.9 $R_{\rm Jup}$} & {9.6 $R_{\rm Jup}$} & {15.3 $R_{\rm Jup}$} & {21.0 $R_{\rm Jup}$} & {26.9 $R_{\rm Jup}$}\\ \hline 
      0.1 $M_{\odot}$ & \cellcolor{gray!35}{0.54} & \cellcolor{gray!35}{0.88} & \cellcolor{gray!35}{1.4} & \cellcolor{gray!35}{1.9} & \cellcolor{gray!35}{2.5} \\ \hline
      0.2 $M_{\odot}$ & {0.32} & \cellcolor{gray!35}{0.52} & \cellcolor{gray!35}{0.82} & \cellcolor{gray!35}{1.1} & \cellcolor{gray!35}{1.4} \\ \hline
      0.3 $M_{\odot}$ & {0.25} & {0.41} & \cellcolor{gray!35}{0.66} & \cellcolor{gray!35}{0.91} & \cellcolor{gray!35}{1.2} \\ \hline
      0.4 $M_{\odot}$ & {0.20} & {0.33} & \cellcolor{gray!35}{0.53} & \cellcolor{gray!35}{0.72} & \cellcolor{gray!35}{0.93} \\ \hline
      0.5 $M_{\odot}$ & {0.16} & {0.26} & {0.41} & \cellcolor{gray!35}{0.57} & \cellcolor{gray!35}{0.72}\\ \hline
      0.6 $M_{\odot}$ & {0.12} & {0.20} & {0.31} & {0.43} & \cellcolor{gray!35}{0.55} \\ \hline
    \end{tabular} \\ [2ex]
    \begin{tabular}{|c|c|c|c|c|c|}
      \hline 
      \multicolumn{6}{|c|}{\textbf{Planet Orbit in Center of HZ}}\\ \hline 
      \multirow{2}{*}{Star Mass} & \multicolumn{5}{|c|}{Moon Semi-major Axis}\\ \cline{2-6}
      & {5.9 $R_{\rm Jup}$} & {9.6 $R_{\rm Jup}$} & {15.3 $R_{\rm Jup}$} & {21.0 $R_{\rm Jup}$} & {26.9 $R_{\rm Jup}$}\\ \hline 
      0.1 $M_{\odot}$ & {0.38} & \cellcolor{gray!35}{0.61} & \cellcolor{gray!35}{0.98} & \cellcolor{gray!35}{1.3} & \cellcolor{gray!35}{1.7}\\ \hline
      0.2 $M_{\odot}$ & {0.22} & {0.36} & \cellcolor{gray!35}{0.57} & \cellcolor{gray!35}{0.79} & \cellcolor{gray!35}{1.0} \\ \hline
      0.3 $M_{\odot}$ &  {0.18} & {0.29} & \cellcolor{gray!35}{0.46} & \cellcolor{gray!35}{0.63} & \cellcolor{gray!35}{0.80} \\ \hline
      0.4 $M_{\odot}$ &  {0.14} & {0.23} & {0.37} & \cellcolor{gray!35}{0.50} & \cellcolor{gray!35}{0.65} \\ \hline
      0.5 $M_{\odot}$ &  {0.11} & {0.18} & {0.29} & {0.39} & \cellcolor{gray!35}{0.51}\\ \hline
      0.6 $M_{\odot}$ &  {0.09} & {0.14} & {0.22} & {0.30} & {0.39}\\ \hline
    \end{tabular}
  \end{center}
\end{table}

\begin{table}
  \caption{3-body stability summary for a Saturn-like host planet. The moon semi-major axes are presented as fractions of $R_{\rm Hill}$. Gray shaded cells represent unstable moon orbits.}
  \label{tab:stab3brhillsat}
  \begin{center}
    \vspace{2 mm}
    \begin{tabular}{|c|c|c|c|c|c|}
      \hline 
      \multicolumn{6}{|c|}{\textbf{Earth-Equivalent Planet Orbit}}\\ \hline 
      \multirow{2}{*}{Star Mass} & \multicolumn{5}{|c|}{Moon Semi-major Axis}\\ \cline{2-6}
      & {5.9 $R_{\rm Sat}$} & {9.6 $R_{\rm Sat}$} & {15.3 $R_{\rm Sat}$} & {21.0 $R_{\rm Sat}$} & {26.9 $R_{\rm Sat}$}\\ \hline 
      0.1 $M_{\odot}$ & \cellcolor{gray!35}{0.66} & \cellcolor{gray!35}{1.1} & \cellcolor{gray!35}{1.7} & \cellcolor{gray!35}{2.3} & \cellcolor{gray!35}{3.0}\\ \hline
      0.2 $M_{\odot}$ & {0.39} & \cellcolor{gray!35}{0.63} & \cellcolor{gray!35}{1.0} & \cellcolor{gray!35}{1.4} & \cellcolor{gray!35}{1.8}\\ \hline
      0.3 $M_{\odot}$ & {0.31} & \cellcolor{gray!35}{0.50} & \cellcolor{gray!35}{0.80} & \cellcolor{gray!35}{1.1} & \cellcolor{gray!35}{1.4}\\ \hline
      0.4 $M_{\odot}$ & {0.25} & {0.40} & \cellcolor{gray!35}{0.64} & \cellcolor{gray!35}{0.88} & \cellcolor{gray!35}{1.1}\\ \hline
      0.5 $M_{\odot}$ & {0.19} & {0.32} & \cellcolor{gray!35}{0.50} & \cellcolor{gray!35}{0.69} & \cellcolor{gray!35}{0.88}\\ \hline
      0.6 $M_{\odot}$ & {0.15} & {0.24} & {0.38} & \cellcolor{gray!35}{0.53} & \cellcolor{gray!35}{0.67}\\ \hline
    \end{tabular} \\ [2ex]
    \begin{tabular}{|c|c|c|c|c|c|}
      \hline 
      \multicolumn{6}{|c|}{\textbf{Planet Orbit in Center of HZ}}\\ \hline 
      \multirow{2}{*}{Star Mass} & \multicolumn{5}{|c|}{Moon Semi-major Axis}\\ \cline{2-6}
      & {5.9 $R_{\rm Sat}$} & {9.6 $R_{\rm Sat}$} & {15.3 $R_{\rm Sat}$} & {21.0 $R_{\rm Sat}$} & {26.9 $R_{\rm Sat}$}\\ \hline 
      0.1 $M_{\odot}$ & \cellcolor{gray!35}{0.46} & \cellcolor{gray!35}{0.75} & \cellcolor{gray!35}{1.2} & \cellcolor{gray!35}{1.6} & \cellcolor{gray!35}{2.1}\\ \hline
      0.2 $M_{\odot}$ & {0.27} & \cellcolor{gray!35}{0.44} & \cellcolor{gray!35}{0.70} & \cellcolor{gray!35}{0.96} & \cellcolor{gray!35}{1.2}\\ \hline
      0.3 $M_{\odot}$ & {0.22} & {0.35} & \cellcolor{gray!35}{0.56} & \cellcolor{gray!35}{0.77} & \cellcolor{gray!35}{0.98}\\ \hline
      0.4 $M_{\odot}$ & {0.17} & {0.28} & \cellcolor{gray!35}{0.45} & \cellcolor{gray!35}{0.61} & \cellcolor{gray!35}{0.79}\\ \hline
      0.5 $M_{\odot}$ & {0.14} & {0.22} & {0.35} & \cellcolor{gray!35}{0.48} & \cellcolor{gray!35}{0.62}\\ \hline
      0.6 $M_{\odot}$ & {0.10} & {0.17} & {0.27} & {0.37} & \cellcolor{gray!35}{0.47}\\ \hline
    \end{tabular}
  \end{center}
\end{table}

With a central star ranging in mass from $0.075\ M_{\odot}$ to $0.6\ M_{\odot}$ and our confined planetary orbits within the relatively close stellar HZs, the exomoon Hill stability became very limited. We defined the stability region by the Hill radius ($R_{\rm Hill}$), given by the relation

\begin{equation} \label{rhill} R_{\rm Hill} = a_{\rm P} \left(\frac{M_{\rm p}}{3M_\star}\right)^{1/3}, \end{equation}

where $a_{\rm P}$ is the semimajor axis of the planet's orbit around the star, $M_{\rm p}$ and $M_\star$ are the masses of the planet and star, respectively. 
An extended study of moon orbital stability was conducted by \cite{domingos06}, who showed that the actual stability region depends upon the eccentricity and orientation of the moon's orbit. For prograde moons the stable region is:

\begin{equation}\label{pro_stable}
a_{\rm s,max} = 0.4895R_{\rm Hill} (1.0000 - 1.0305\,e_{\rm p} - 0.2738e_{\rm s}),
\end{equation}

where $e_{\rm p}$ and $e_{\rm s}$ are the orbital eccentricities of the planet and satellite, respectively. Following Eqs.~(\ref{rhill}) and (\ref{pro_stable}), we expected a stability limit of $\sim0.4\,R_{\rm Hill}$ for our chosen moon orbits, and so we only simulated systems for which the moon's semi-major axis about the planet was $<0.5\,R_{\rm Hill}$. Our initial tests showed that for even the tightest exomoon orbit, which was an Io-like orbit at $a = 5.9\,R_{\rm p}$, no stable systems were found around stars less than 0.1\,$M_{\odot}$. We therefore limited extended simulations to stellar masses $\geq 0.1\ M_{\odot}$, using intervals of $0.1\,M_{\odot}$ for our systems (i.e., the considered star masses were 0.1, 0.2, 0.3, 0.4, 0.5, and 0.6\,$M_{\odot}$).

The stable region around a planet slowly increased with stellar mass (i.e. with increasingly larger HZ distances). As explained in Sect.~\ref{moon_model}, we considered a discrete range of moon orbital distances for each star-planet pair. Not surprising, many of the wider moon orbits were unstable for the lowest star masses. This instability resulted in the moon becoming unbound from the planet. We found that the 3-body simulations were only able to maintain long-term stability for a satellite orbit of $a_{\rm s}$ $\lesssim$ $0.4\ R_{\rm Hill}$, which was the predicted value described above. A summary of the stable 3-body systems is provided in Tables~\ref{tab:stab3brhilljup} and \ref{tab:stab3brhillsat}. Note in the tables that the moon semi-major axes are modeled after those of the Solar System moons Io, Europa, Ganymede, Titan, and Callisto, respectively. The planet distances from the star represented in the tables range from about 0.05\,AU for the lowest stellar mass to about 0.4\,AU for the highest stellar mass.

\textcolor{black}{Due to perturbations from the third body the eccentricity of the moon experienced short-term oscillations with periods relating to the moon's orbital period around the planet and the planet's orbital period around the star. Because of this, when reporting orbital parameters and surface heat flux for the moon the values were first averaged over the moon's orbit and then averaged again over multiple planet orbits.}

\subsection{Extended Simulation Results} \label{3bresults}

After completing our investigations into system stability and short-term behavior, we were left with a set of 3-body systems for which we could expect stability and reasonably predict the long-term behavior. \textcolor{black}{In other words, our systems were feasable from a strictly dynamical perspective.} The integrated times for long-term simulations were extended until one of three results was achieved: (1) the average surface heat flux fell well below the proposed maximum limit for habitability, $h_{\rm max}=2\,{\rm W\,m}^{-2}$; (2) the average surface heat flux settled to a reasonably constant value; or (3) the tidal evolution was slow enough that it was not practical to continue the simulation further.

\begin{figure*}
     \begin{center}
       \subfigure[]{%
            \label{fig:cenj1io}
           \includegraphics[width=0.5\textwidth]{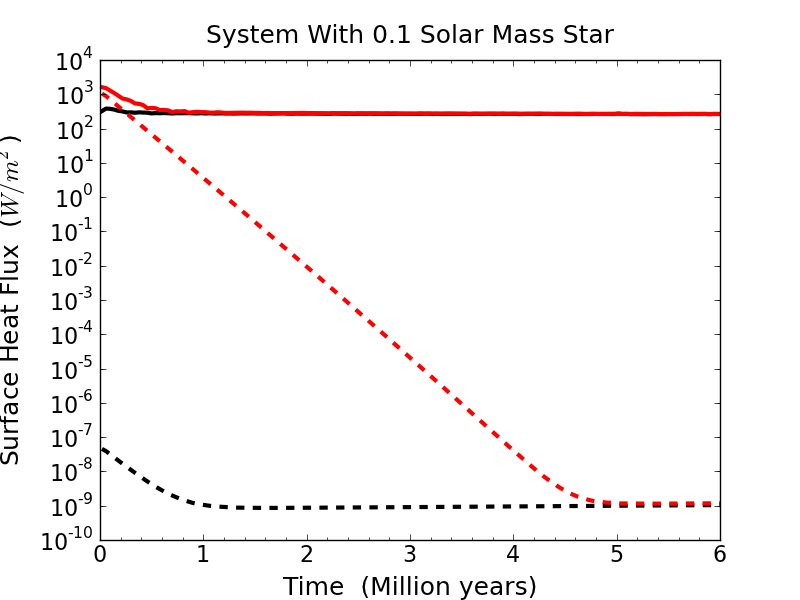}
        }%
        \subfigure[]{%
            \label{fig:cenj6io}
           \includegraphics[width=0.5\textwidth]{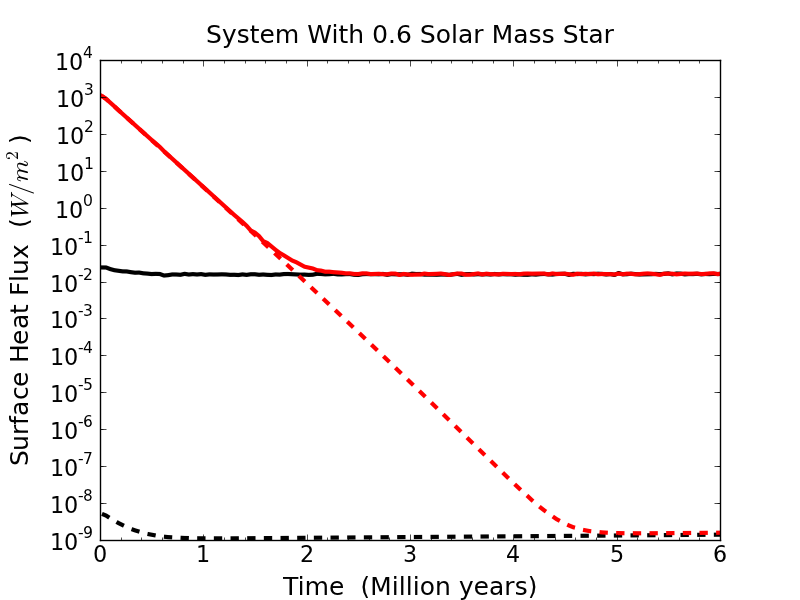}
        }%

        \includegraphics[width=.7\textwidth]{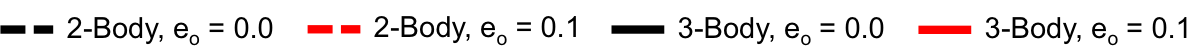}
    \end{center}
    \caption{Examples of the orbital evolution of a Mars-mass moon in an Io-like orbit around a Jupiter-like planet.}
   \label{fig:cenjio}
\end{figure*}

\begin{figure*}
     \begin{center}
        \subfigure[]{%
            \label{fig:cenj2eur}
            \includegraphics[width=0.5\textwidth]{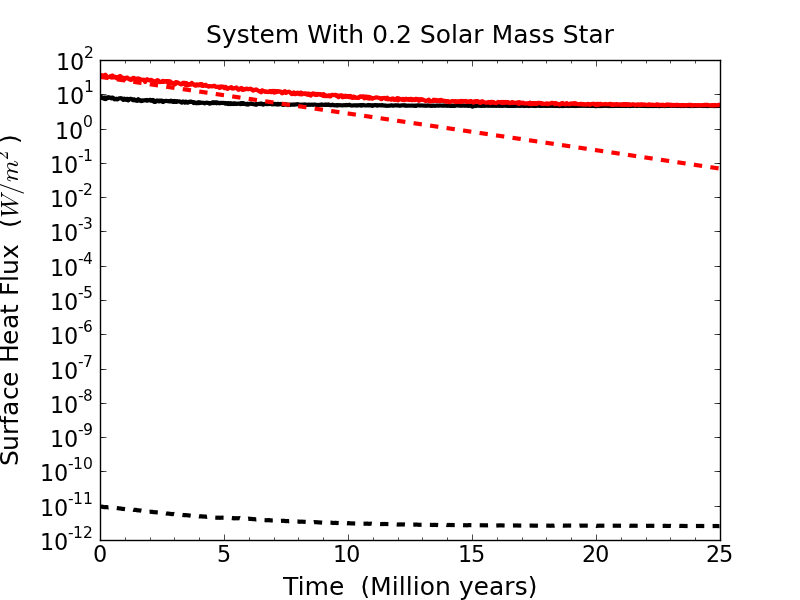}
        }%
        \subfigure[]{%
            \label{fig:cenj6eur}
           \includegraphics[width=0.5\textwidth]{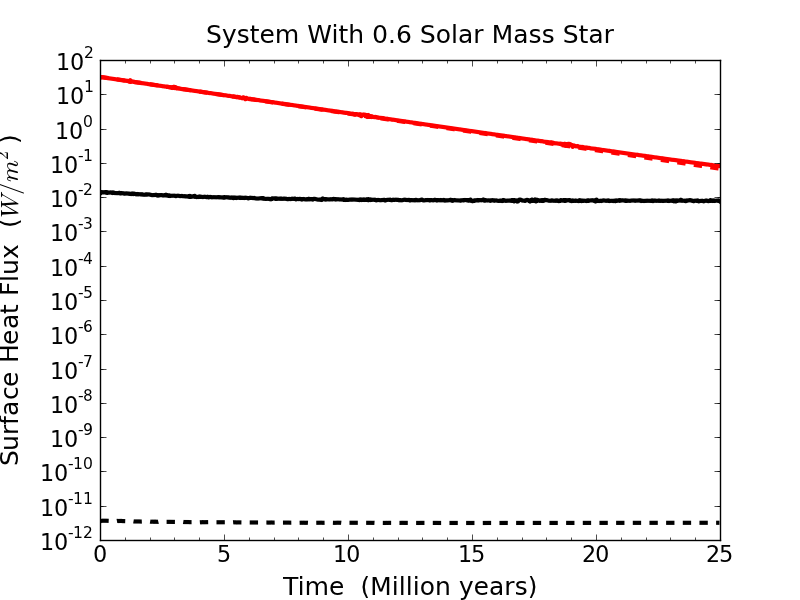}
        }%

        \includegraphics[width=.7\textwidth]{fig4567legend.png}
    \end{center}
    \caption{Examples of the orbital evolution of a Mars-mass moon in a Europa-like orbit around a Jupiter-like planet.}
   \label{fig:cenjeur}
\end{figure*}

\begin{figure*}
     \begin{center}
        \subfigure[]{%
            \label{fig:cens2io}
           \includegraphics[width=0.5\textwidth]{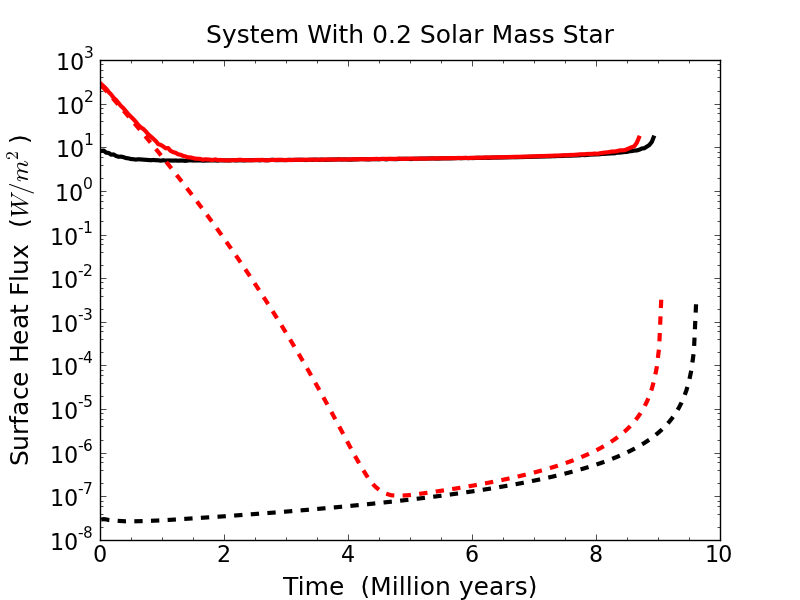}
        }%
        \subfigure[]{%
            \label{fig:cens6io}
           \includegraphics[width=0.5\textwidth]{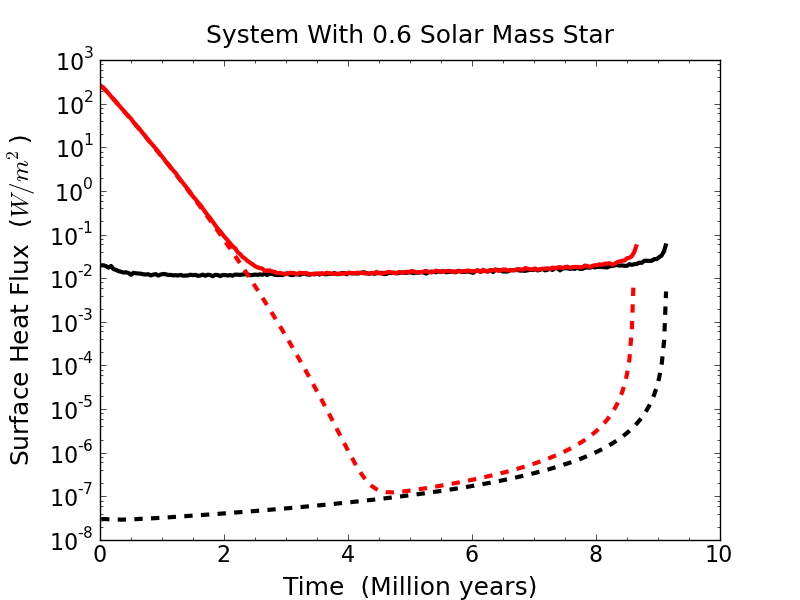}
        }%

        \includegraphics[width=.7\textwidth]{fig4567legend.png}
    \end{center}
    \caption{Examples of the orbital evolution of a Mars-mass moon in an Io-like orbit around a Saturn-like planet.}
   \label{fig:censio}
\end{figure*}

\begin{figure*}
     \begin{center}
        \subfigure[]{%
            \label{fig:cens3eur}
           \includegraphics[width=0.5\textwidth]{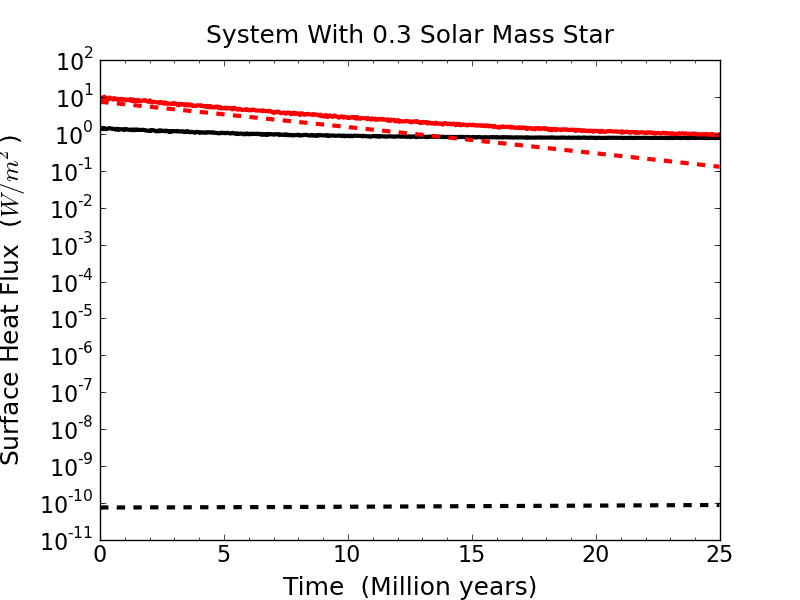}
        }%
        \subfigure[]{%
            \label{fig:cens6eur}
           \includegraphics[width=0.5\textwidth]{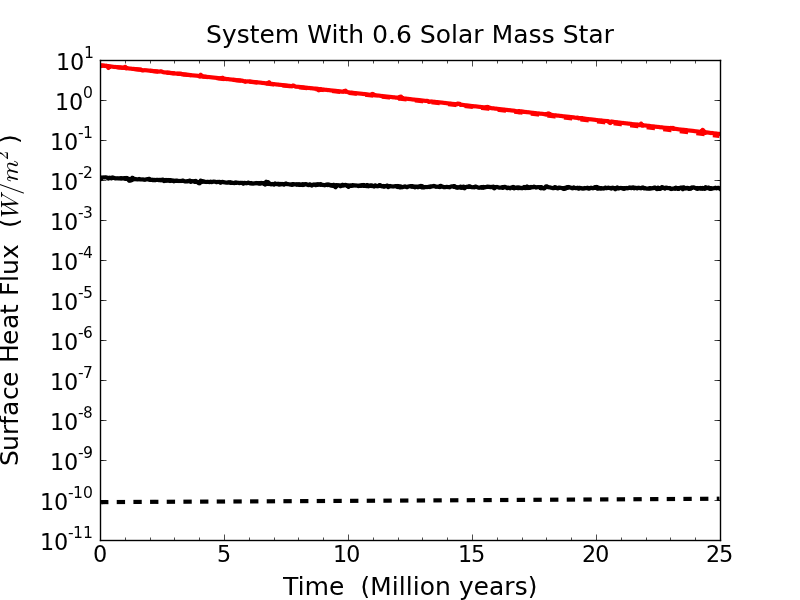}
        }%

        \includegraphics[width=.7\textwidth]{fig4567legend.png}
    \end{center}
    \caption{Examples of the orbital evolution of a Mars-mass moon in a Europa-like orbit around a Saturn-like planet.}
   \label{fig:censeur}
\end{figure*}

Examples of tidal evolution plots for Io-like and Europa-like moons with a Jupiter-like host planet are shown in Figs.~\ref{fig:cenjio} and \ref{fig:cenjeur}. Similar plots for a Saturn-like host planet are included in Figs.~\ref{fig:censio} and \ref{fig:censeur}. Plots for Ganymede-like, Callisto-like, and Titan-like moon simulations are not included since there was little to no change in the surface heat flux of those moons during the integrated time periods. Displayed in these figures is the average surface heat flux ($h$) in the moon as a function of time, with $h$ defined by Eq.~(\ref{surfheat}). The individual plots in each figure represent low and high limit mass values considered for the central star. Note that a greater star mass signifies a larger orbital distance for the planet. Each plot includes four curves with two solid ones referring to 3-body systems and two dashed ones referring to 2-body systems. Both pair of curves contains one case with zero initial eccentricity and another scenario with an initial eccentricity of 0.1. Comparing the 2-body and 3-body curves is useful in demonstrating the star's influence on the long-term tidal evolution of the moon.

\subsection{Discussion} \label{3bdiscuss}

The focus of this study involves the long-term behavior of the solid red curves in Figs.~\ref{fig:cenjio} through \ref{fig:censeur}. With them we test the extent to which a low-mass star has influence over a moon's evolution in the HZ, as a consequence of the short distance of the HZ. This is done by comparing the solid red curve to the dashed red curve, which represents an isolated planet-moon binary that began the simulation with the same initial conditions, minus the star. The dashed red curve shows what the solid red curve would look like without the influence of the central star.

Since tidal heating works to circularize the orbits, initially circular orbits (black curves) should continue with little to no heating, unless there are significant perturbations to the orbit. The purpose of these simulations was to test if the central star would excite the heating to the same steady state value that was achieved through the slow decay of an initially higher value. If that result was observed, then there is confidence that a non-zero, steady-state value for the surface heat flux was the product of perturbations from the star, and not simply based on our choice of initial conditions or an external artifact of our computer code. This anticipated result was indeed observed for Io-like and Europa-like orbital distances, but not for the wider orbits (Ganymede-like, Titan-like, and Callisto-like orbits). This difference between the eccentric and circular simulations for wide orbit systems remained constant throughout a test period of 10\,Myr. 

Our specific interest relates to the difference in the solid red and dashed red curves by the end of each simulation. When the slope of the solid red line deviates from the dashed red line and eventually levels off, it suggests a sustainable value for the surface heat flux. This indicates that the star has significant influence on the moon, and the continual perturbations are restricting the moon's eccentricity from reaching zero. This behavior is observed for all the Io-like and Europa-like orbital distances. For our systems the timescales necessary to achieve the constant state were more than double the estimates listed in Table \ref{tab:dampscales}. For convenience, we will refer to Io-like and Europa like orbits as ``short'' orbits.  At these distances the initially circular simulations eventually reached the same approximate steady state values as the non-circular systems. This result helps to confirm the degree of excitation to which the central star can influence the tidal evolution of a moon. 

For the shorter orbits, our results show that perturbations from a low-mass star can prevent circularization and are able to maintain tidal heating in the moon for extended periods of time. While this effect was predicted \citep{heller12}, this is the first instance in which it has actually been tested with an evolution model that simultaneously and self-consistently considered gravitational and tidal effects.

At orbital distances $a_{\rm s} \gtrsim 15 R_{\rm p}$, little change to the surface heat flux occurred over 10\,Myr. For convenience, reference to these distances as ``wide'' orbits will correspond to Ganymede, Titan, and Callisto-like moon orbits. In wide orbits, we found little deviation from the 2-body models, which was expected based on estimates of eccentricity damping timescales and early 2-body test runs. Tidal heat fluxes for these orbits are $\lesssim0.1\,{\rm W\,m}^{-2}$, even with the influence of the star. This is an order of magnitude less than the shorter orbit moons and well below $h_{\rm max}$. Such low heating rates would not be expected to noticeably affect the orbit over the timescales considered. 

An examination of Fig.~\ref{fig:censio} shows unique behavior that is not observed in the other figures. In these simulations the surface heat flux rapidly increases near the end of the simulated time. The cause for this behavior was already discussed in Sect.~\ref{tvsnot}. Using 2-body models, we found that Mars-like moons with Io-like orbits and Saturn-like host planets experience significant evolution in their semi-major axis as a result of tidal torques and exchanges of angular momentum. Our 3-body results indicate that the star's influence does little to change this long-term behavior. In this case, the rapid rise in tidal heating is due to the equally rapid decay of the moon's semi-major axis as the moon spirals ever closer to the planet. The sudden end to the plots in Fig.~\ref{fig:censio} reflects the early termination of the simulations when the moon approached the Roche radius at $2\,R_{\rm p}$, where it would experience tidal disruption. Note that the complete inward spiral occurred in $<10$\,Myr.

\subsubsection{A Complete Simulation Summary} \label{enddis}

Our results for the 3-body simulations of Io-like and Europa-like satellite systems in the center of the HZ are summarized in Table \ref{tab:cenend}, while the results for systems at Earth-equivalent distances are summarized in Table \ref{tab:eqend}. Our interest is in heating values which are developed after noticeable tidal evolution and may be maintained for extended periods of time by stellar perturbations. The tables do not include results for wider orbits since, as discussed previously, there was no significant orbital evolution in these systems during the considered time periods. Therefore, surface heating values for wide orbit moons would simply reflect our choice for the initial orbital parameters of those systems and would not allow for any conclusive statements as to their ultimate tidal evolution. It is worth noting that very few wide orbit moons are even stable in \textcolor{black}{our low-mass} star range, and of those, all wide orbit systems had surface heat flux values below $2\,{\rm W\,m}^{-2}$ for moon eccentricities at or below 0.1.  

\begin{table*}
  \caption{3-body moon evolution summary for systems in the center of the stellar HZ. Values represent the average at the end of each simulation. Shaded rows are 2-body planet/moon systems. Red text indicates steady state values above $h_{\rm max}$. Values of ** indicate the moon spiraled into the planet.}
  \label{tab:cenend}
  \begin{center}
    \begin{tabular}{|c|c|c|c|c|c|c|c|c|} 
      \hline 
      \multicolumn{9}{ |c| }{Jupiter-like Host Planet} \\ \hline 
      Moon Distance & Star Mass & Planet & $h$ & $h_{\rm conv}$ & $e$ & $a/a_0$ & $F_{\rm glob}$ & Sim. Time\\ 
      ($R_{\rm Jup}$) & ($M_{\odot}$) & Distance ($AU$) & (W/m$^2$) & (W/m$^2$) & {} & {} & (W/m$^2$) & (10$^6$ Years)\\ \hline 
      \multirow{7}{*}{\parbox{2.5cm}{\centering Io-like\\5.9}} & \cellcolor{blue!15}{$\mathbf{-}$} & \cellcolor{blue!15}{$\mathbf{-}$} & \cellcolor{blue!15}{0.00} & \cellcolor{blue!15}{0.00} & \cellcolor{blue!15}{0.000} & \cellcolor{blue!15}{0.95} & \cellcolor{blue!15}{$\mathbf{-}$} & \cellcolor{blue!15}{10}\\ \cline{2-9} 
       & 0.1 & 0.051 & \textcolor{red}{265} & \textcolor{red}{422} & 0.030 & 0.82 & 371 & 10\\ \cline{2-9} 
       & 0.2 & 0.11 & \textcolor{red}{6.46} & \textcolor{red}{11.1} & 0.008 & 0.94 & 113 & 10\\ \cline{2-9} 
       & 0.3 & 0.16 & 1.51 & \textcolor{red}{2.59} & 0.004 & 0.95 & 107 & 10\\ \cline{2-9} 
       & 0.4 & 0.22 & 0.38 & 0.66 & 0.002 & 0.95 & 106 & 10\\ \cline{2-9} 
       & 0.5 & 0.30 & 0.08 & 0.14 & 0.001 & 0.95 & 107 & 10\\ \cline{2-9} 
       & 0.6 & 0.41 & 0.02 & 0.03 & 0.000 & 0.95 & 109 & 10\\ \hline 
      \multirow{6}{*}{\parbox{2.5cm}{\centering Europa-like\\9.6}} & \cellcolor{blue!15}{$\mathbf{-}$} & \cellcolor{blue!15}{$\mathbf{-}$} & \cellcolor{blue!15}{0.07} & \cellcolor{blue!15}{0.07} & \cellcolor{blue!15}{0.005} & \cellcolor{blue!15}{0.99} & \cellcolor{blue!15}{$\mathbf{-}$} & \cellcolor{blue!15}{25}\\ \cline{2-9} 
       & 0.2 & 0.11 & \textcolor{red}{4.94} & \textcolor{red}{7.22} & 0.049 & 0.98 & 109 & 25\\ \cline{2-8} 
       & 0.3 & 0.16 & 0.97 & 1.53 & 0.022 & 0.99 & 105 & 25\\ \cline{2-9} 
       & 0.4 & 0.22 & 0.29 & 0.39 & 0.011 & 0.99 & 104 & 25\\ \cline{2-9} 
       & 0.5 & 0.30 & 0.12 & 0.12 & 0.006 & 0.99 & 105 & 25\\ \cline{2-9} 
       & 0.6 & 0.41 & 0.08 & 0.08 & 0.005 & 0.99 & 108 & 25\\ \hline 
    \end{tabular}

    \vspace{5mm}

    \begin{tabular}{|c|c|c|c|c|c|c|c|c|} 
      \hline 
      \multicolumn{9}{ |c| }{Saturn-like Host Planet} \\ \hline 
      Moon Distance & Star Mass & Planet & $h$ & $h_{\rm conv}$ & $e$ & $a/a_0$ & $F_{\rm glob}$ & Sim. Time\\ 
      ($R_{\rm Sat}$) & ($M_{\odot}$) & Distance ($AU$)& (W/m$^2$) & (W/m$^2$) & {} & {} & (W/m$^2$) & (10$^6$ Years)\\ \hline 
      \multirow{6}{*}{\parbox{2.5cm}{\centering Io-like\\5.9}} & \cellcolor{blue!15}{$\mathbf{-}$} & \cellcolor{blue!15}{$\mathbf{-}$} & \cellcolor{blue!15}{**} & \cellcolor{blue!15}{**} & \cellcolor{blue!15}{**} & \cellcolor{blue!15}{**} & \cellcolor{blue!15}{**} & \cellcolor{blue!15}{10}\\ \cline{2-9} 
       & 0.2 & 0.11 & {**} & {**} & {**} & {**} & {**} & {10}\\ \cline{2-9} 
       & 0.3 & 0.16 & {**} & {**} & {**} & {**} & {**} & {10}\\ \cline{2-9} 
       & 0.4 & 0.22 & {**} & {**} & {**} & {**} & {**} & {10}\\ \cline{2-9} 
       & 0.5 & 0.30 & {**} & {**} & {**} & {**} & {**} & {10}\\ \cline{2-9} 
       & 0.6 & 0.41 & {**} & {**} & {**} & {**} & {**} & {10}\\ \hline 
      \multirow{5}{*}{\parbox{2.5cm}{\centering Europa-like\\9.6}} & \cellcolor{blue!15}{$\mathbf{-}$} & \cellcolor{blue!15}{$\mathbf{-}$} & \cellcolor{blue!15}{0.13} & \cellcolor{blue!15}{0.13} & \cellcolor{blue!15}{0.012} & \cellcolor{blue!15}{0.97} & \cellcolor{blue!15}{$\mathbf{-}$} & \cellcolor{blue!15}{25}\\ \cline{2-8} 
       & 0.3 & 0.16 & 0.97 & 1.38 & 0.041 & 0.97 & 105 & 25\\ \cline{2-9} 
       & 0.4 & 0.22 & 0.29 & 0.36 & 0.020 & 0.97 & 104 & 25\\ \cline{2-9} 
       & 0.5 & 0.30 & 0.15 & 0.15 & 0.013 & 0.97 & 106 & 25\\ \cline{2-9} 
       & 0.6 & 0.41 & 0.14 & 0.14 & 0.013 & 0.97 & 108 & 25\\ \hline 
    \end{tabular}
  \end{center}
\end{table*}

\begin{table*}
  \caption{3-body moon evolution summary for systems at Earth-equivalent distances. Values represent the average at the end of each simulation. Shaded rows are 2-body planet/moon systems. Red text indicates steady state values above $h_{\rm max}$. Values of ** indicates the moon spiraled into the planet.}
  \label{tab:eqend}
  \begin{center}
    \begin{tabular}{|c|c|c|c|c|c|c|c|c|} 
      \hline 
      \multicolumn{9}{ |c| }{Jupiter-like Host Planet} \\ \hline 
      Moon Distance & Star Mass & Planet & $h$ & $h_{\rm conv}$ & $e$ & $a/a_0$ & $F_{\rm glob}$ & Sim. Time\\ 
      ($R_{\rm Jup}$) & ($M_{\odot}$) & Distance ($AU$) & (W/m$^2$) & (W/m$^2$) & {} & {} & (W/m$^2$) & (10$^6$ Years)\\ \hline 
      \multirow{6}{*}{\parbox{2.5cm}{\centering Io-like\\5.9}} & \cellcolor{blue!15}{$\mathbf{-}$} & \cellcolor{blue!15}{$\mathbf{-}$} & \cellcolor{blue!15}{0.00} & \cellcolor{blue!15}{0.00} & \cellcolor{blue!15}{0.000} & \cellcolor{blue!15}{0.95} & \cellcolor{blue!15}{$\mathbf{-}$} & \cellcolor{blue!15}{10}\\ \cline{2-9} 
       & 0.2 & 0.076 & \textcolor{red}{72.7} & \textcolor{red}{121} & 0.024 & 0.92 & 292 & 10\\ \cline{2-9} 
       & 0.3 & 0.11 & \textcolor{red}{15.4} & \textcolor{red}{26.7} & 0.013 & 0.94 & 234 & 10\\ \cline{2-9} 
       & 0.4 & 0.15 & \textcolor{red}{3.68} & \textcolor{red}{6.45} & 0.006 & 0.94 & 223 & 10\\ \cline{2-9} 
       & 0.5 & 0.21 & 0.76 & 1.35 & 0.003 & 0.95 & 221 & 10\\ \cline{2-9} 
       & 0.6 & 0.29 & 0.14 & 0.25 & 0.001 & 0.95 & 222 & 10\\ \hline 
      \multirow{5}{*}{\parbox{2.5cm}{\centering Europa-like\\9.6}} & \cellcolor{blue!15}{$\mathbf{-}$} &  \cellcolor{blue!15}{$\mathbf{-}$} & \cellcolor{blue!15}{0.07} & \cellcolor{blue!15}{0.07} & \cellcolor{blue!15}{0.005} & \cellcolor{blue!15}{0.99} & \cellcolor{blue!15}{$\mathbf{-}$} & \cellcolor{blue!15}{25}\\ \cline{2-9} 
       & 0.3 & 0.11 & \textcolor{red}{13.3} & \textcolor{red}{19.9} & 0.056 & 0.91 & 229 & 25\\ \cline{2-9} 
       & 0.4 & 0.15 & \textcolor{red}{2.42} & \textcolor{red}{3.94} & 0.035 & 0.98 & 218 & 25\\ \cline{2-9} 
       & 0.5 & 0.21 & 0.48 & 0.77 & 0.015 & 0.99 & 217 & 25\\ \cline{2-9} 
       & 0.6 & 0.29 & 0.15 & 0.18 & 0.008 & 0.99 & 219 & 25\\ \hline 
    \end{tabular}

    \vspace{5mm}

    \begin{tabular}{|c|c|c|c|c|c|c|c|c|} 
      \hline 
      \multicolumn{9}{ |c| }{Saturn-like Host Planet} \\ \hline 
      Moon Distance & Star Mass & Planet & $h$ & $h_{\rm conv}$ & $e$ & $a/a_0$ & $F_{\rm glob}$ & Sim. Time\\ 
      ($R_{\rm Sat}$) & ($M_{\odot}$) & Distance ($AU$) & (W/m$^2$) & (W/m$^2$) & {} & {} & (W/m$^2$) & (10$^6$ Years)\\ \hline 
      \multirow{6}{*}{\parbox{2.5cm}{\centering Io-like\\5.9}} & \cellcolor{blue!15}{$\mathbf{-}$} & \cellcolor{blue!15}{$\mathbf{-}$} &  \cellcolor{blue!15}{**} & \cellcolor{blue!15}{**} & \cellcolor{blue!15}{**} & \cellcolor{blue!15}{**} & \cellcolor{blue!15}{**} & \cellcolor{blue!15}{10}\\ \cline{2-9} 
       & 0.2 & 0.076 & {**} & {**} & {**} & {**} & {**} & {10}\\ \cline{2-9} 
       & 0.3 & 0.11 & {**} & {**} & {**} & {**} & {**} & {10}\\ \cline{2-9} 
       & 0.4 & 0.15 & {**} & {**} & {**} & {**} & {**} & {10}\\ \cline{2-9} 
       & 0.5 & 0.21 & {**} & {**} & {**} & {**} & {**} & {10}\\ \cline{2-9} 
       & 0.6 & 0.29 & {**} & {**} & {**} & {**} & {**} & {10}\\ \hline 
      \multirow{4}{*}{\parbox{2.5cm}{\centering Europa-like\\9.6}} & \cellcolor{blue!15}{$\mathbf{-}$} &  \cellcolor{blue!15}{$\mathbf{-}$} & \cellcolor{blue!15}{0.13} & \cellcolor{blue!15}{0.13} & \cellcolor{blue!15}{0.012} & \cellcolor{blue!15}{0.97} & \cellcolor{blue!15}{$\mathbf{-}$} & \cellcolor{blue!15}{25}\\ \cline{2-9} 
       & 0.4 & 0.15 & \textcolor{red}{2.64} & \textcolor{red}{3.71} & 0.068 & 0.96 & 218 & 25\\ \cline{2-9} 
       & 0.5 & 0.21 & 0.51 & 0.70 & 0.029 & 0.97 & 217 & 25\\ \cline{2-9} 
       & 0.6 & 0.29 & 0.18 & 0.19 & 0.015 & 0.97 & 219 & 25\\ \hline 
    \end{tabular}
  \end{center}
\end{table*}

Tables \ref{tab:cenend} and \ref{tab:eqend} show the average values at the end of each simulation and only include simulations that started with $e_0 = 0.1$ for the moon. Included is the moon's average surface heat flux, the average eccentricity, and the average semi-major axis (in comparison to its initial value). The initial value for the semi-major axis is listed as the moon distance. We also included the `conventional' surface heat flux $h_{\rm conv}$. This value was calculated with Eq.~(\ref{eq:hconv}) using the simulation results for the final eccentricity and semi-major axis and serves as a comparison between our chosen model and the conventional model for tidal heating. In regards to overall exomoon habitability, we included the orbit-averaged global flux ($F_{\rm glob}$) received by a satellite, as defined by Eq.~(\ref{Fglob}). The total integration time considered for each simulation is also included.

Results from a 2-body simulation for each moon orbital distance were also included in Tables \ref{tab:cenend} and \ref{tab:eqend} to demonstrate the influence of the low-mass star in comparison to an isolated planet-moon system. The planet's spin rate in these simulations reflected a tidally locked rotation around a 0.6 solar mass star, which is the slowest of all considered planet spins.

As discussed in the previous subsection, moons with initial Io-like orbits and Saturn-like host planets spiraled into the planet after $\approx10$\,Myr. Over that same time period, distortion torques were much less effective at evolving the semi-major axis for Io-like orbits around Jupiter-like hosts. Note that for a given star mass, the two planets were at equal distances from the star and they both started with a tidally locked rotation relative to the star. This results in approximately equal spin rates for the two planets. The relative effectiveness of the torques is then due to the difference in orbital distances for the satellites. While moons with a Jupiter host evolved much slower in comparison to a Saturn host, their inward migration was still noticeable. If we assume the inward migration rate will only increase, we estimate a total lifetime $<200$\,Myr for a Mars-like moon with an initial Io-like orbit and Jupiter-like host planet in the HZ of a \textcolor{black}{low-mass} star.

\begin{figure*}
     \begin{center}
       $\begin{array}{c}
            \includegraphics[width=0.5\textwidth]{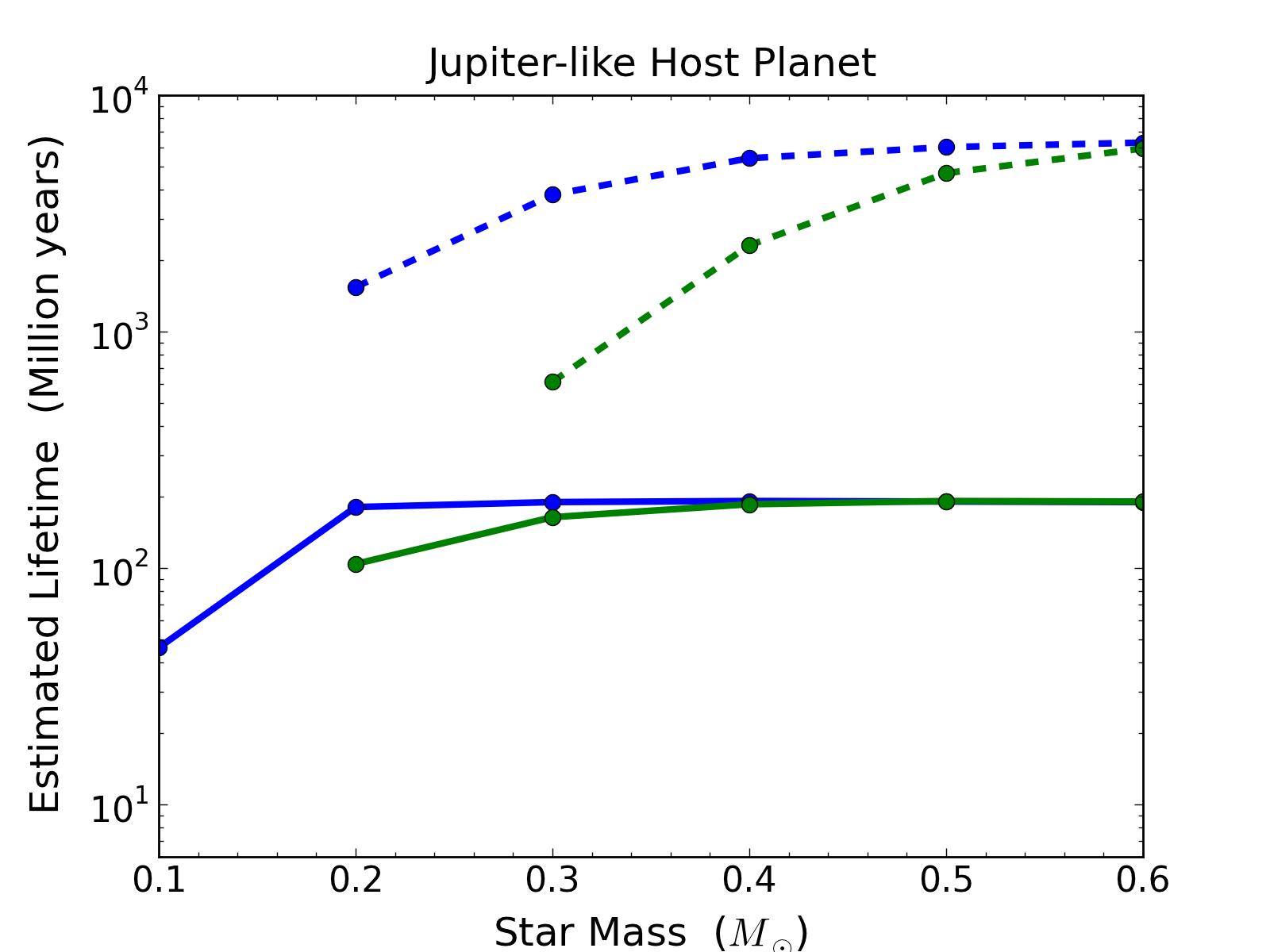}
            \includegraphics[width=0.5\textwidth]{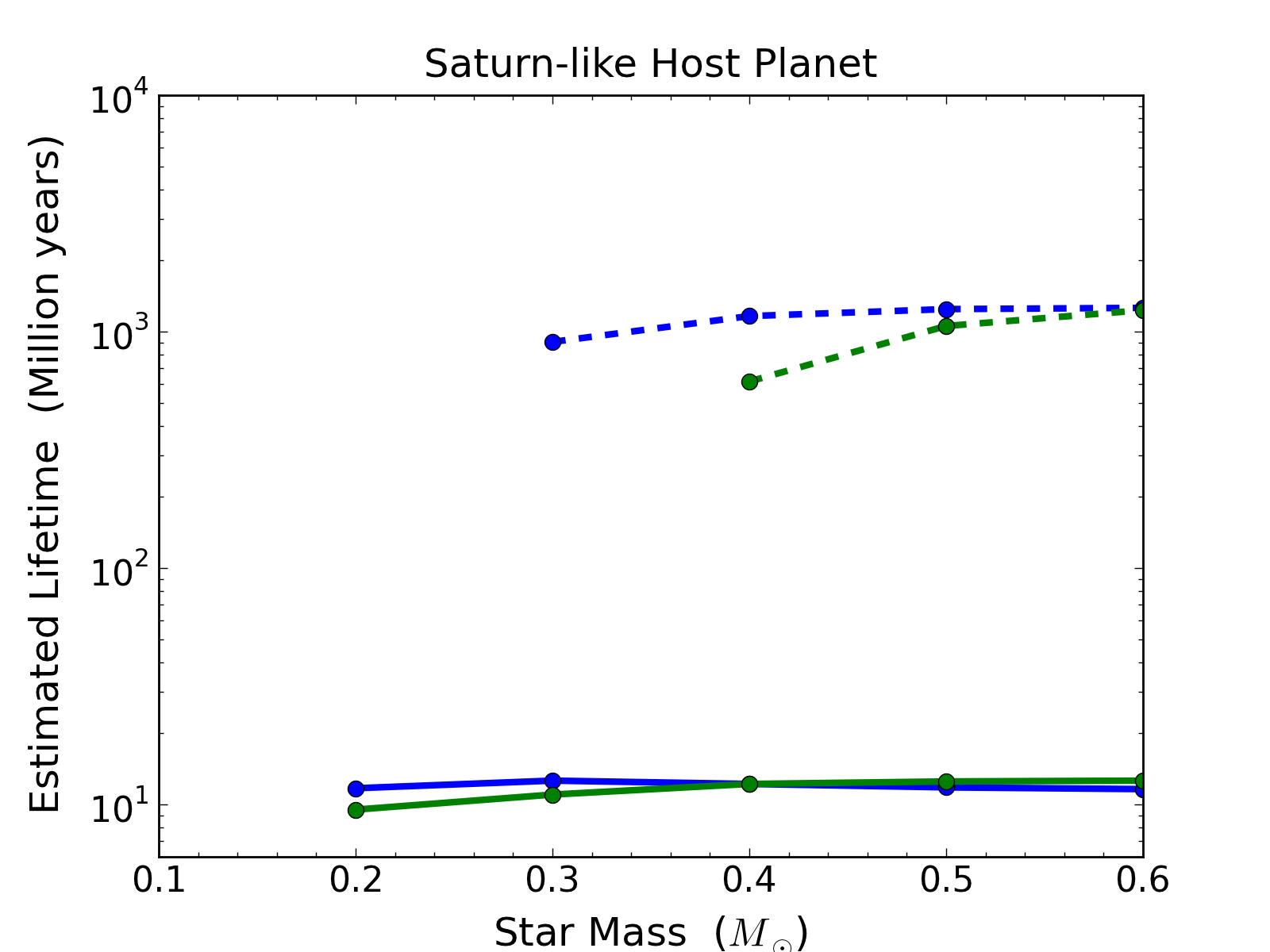}
          \end{array}$
    \includegraphics[width=.8\textwidth]{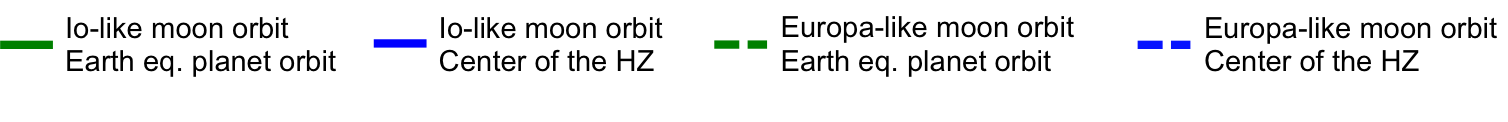}
    \end{center}
    \caption{Estimated lifetimes from each 3-body simulation for the moon to completely spiral into the host planet.}
   \label{fig:lifetimes}
\end{figure*}

Lifetime estimates for the inward spiral of all the Io-like and Europa-like moon orbit simulations are shown in Fig.~\ref{fig:lifetimes}, which shows results from simulations that started with circular moon orbits and thereby limits orbital effects of tidal heating and emphasizes evolution due to distortion torques. Simulations that started with eccentric orbits have shorter estimates for moon lifetimes. Estimated lifetimes for Europa-like orbits are significantly larger than those for Io-like orbits, around a maximum of 1\,Gyr for Saturn-like host planets and several Gyr for a Jupiter-like host. Figure~\ref{fig:lifetimes} indicates that the lifetime of moons at larger orbital distances are more sensitive to the planet's position in the HZ since they are more weakly bound to the planet and can experience greater influence from the central star. The sensitivity, however, becomes negligible for HZ distances around 0.6 solar mass stars. 

One result that stands out in Tables \ref{tab:cenend} and \ref{tab:eqend} is the difference between the conventional model for surface heat flux and our chosen model.  Slightly lower conventional values apply to systems involving higher eccentricities. As mentioned in Sect.~\ref{tid_theory}, the conventional model can break down for high eccentricities, while our chosen model is still appropriate at large values. For systems involving lower eccentricities, $h_{\rm conv}$ is consistently higher than $h$. The difference in the two values scales down for the lowest eccentricities until there is little difference between them. The greatest total difference was less than a factor of 2. Since the exact mechanisms of tidal dissipation are still poorly understood, some difference between two separate tidal models is not surprising.

A similarity between the two tidal models is a direct dependence on the tidal Love number $k_{\rm L}$ and an inverse dependence on the dissipation factor $Q$. Higher values for $Q$ decrease the heating estimates and represent lower dissipation rates in the moon. Higher values for $k_{\rm L}$ produce higher estimates for tidal heating and represent an increased susceptibility of the moon's shape to change in response to a tidal potential. This would also produce a larger acceleration due to the quadrupole moment of the moon, resulting in a shorter inward spiral towards the planet. Recent measurements of these two parameters for the solar system planet Mars contain significant uncertainty \citep[e.g. $k_{\rm L} = 0.148 \pm 0.017;\ Q = 88 \pm 16$;][]{nimmo13}. While these uncertainties can be considered in the simulation results for a single orbit, the long term effects of varying $k_{\rm L}$ and $Q$ would require each simulation to be repeated with the different values. 

\begin{figure*}
     \centering
	  $\begin{array}{c}
    		\includegraphics[width=.49\textwidth]{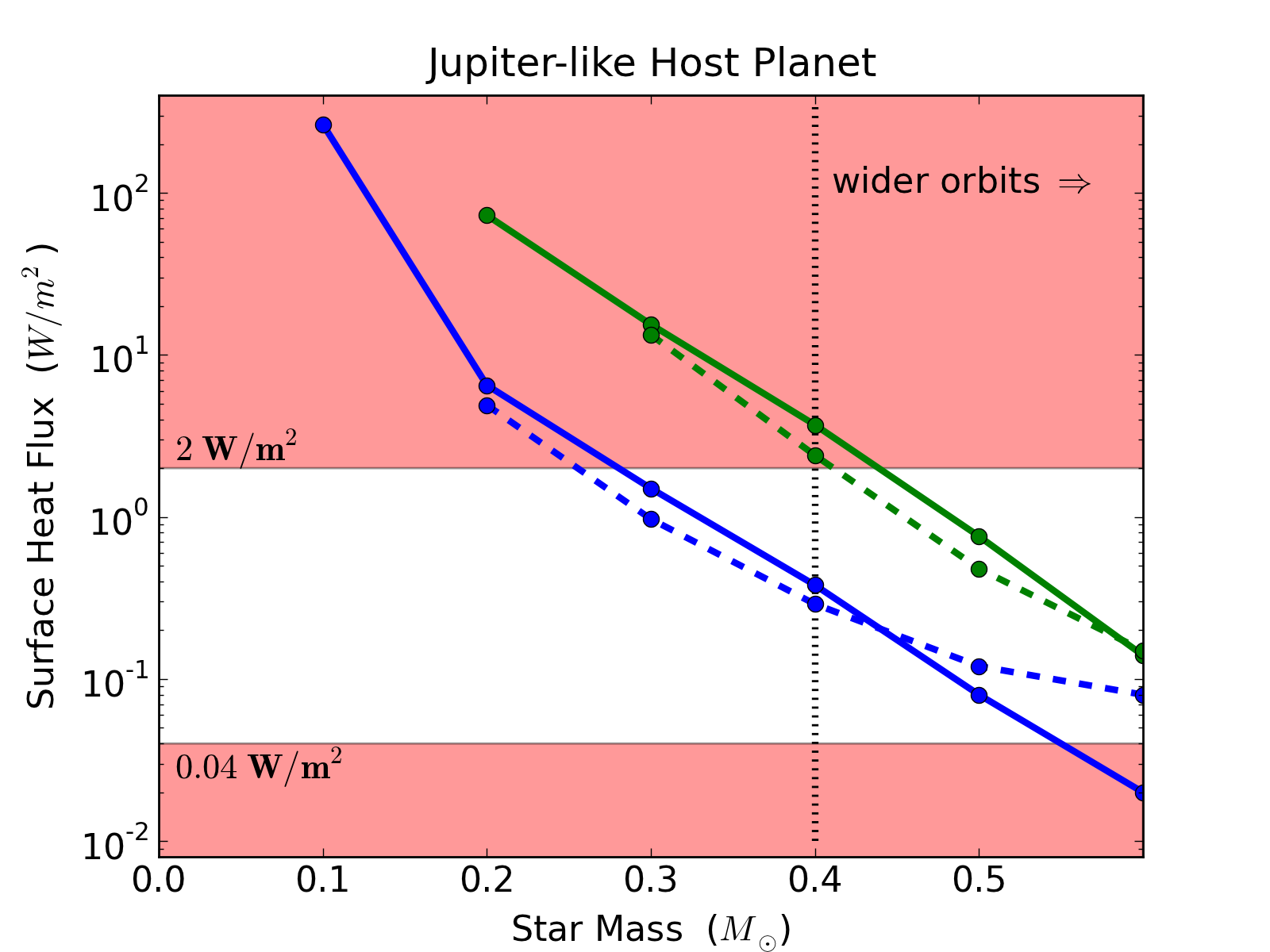}    
    		\includegraphics[width=.49\textwidth]{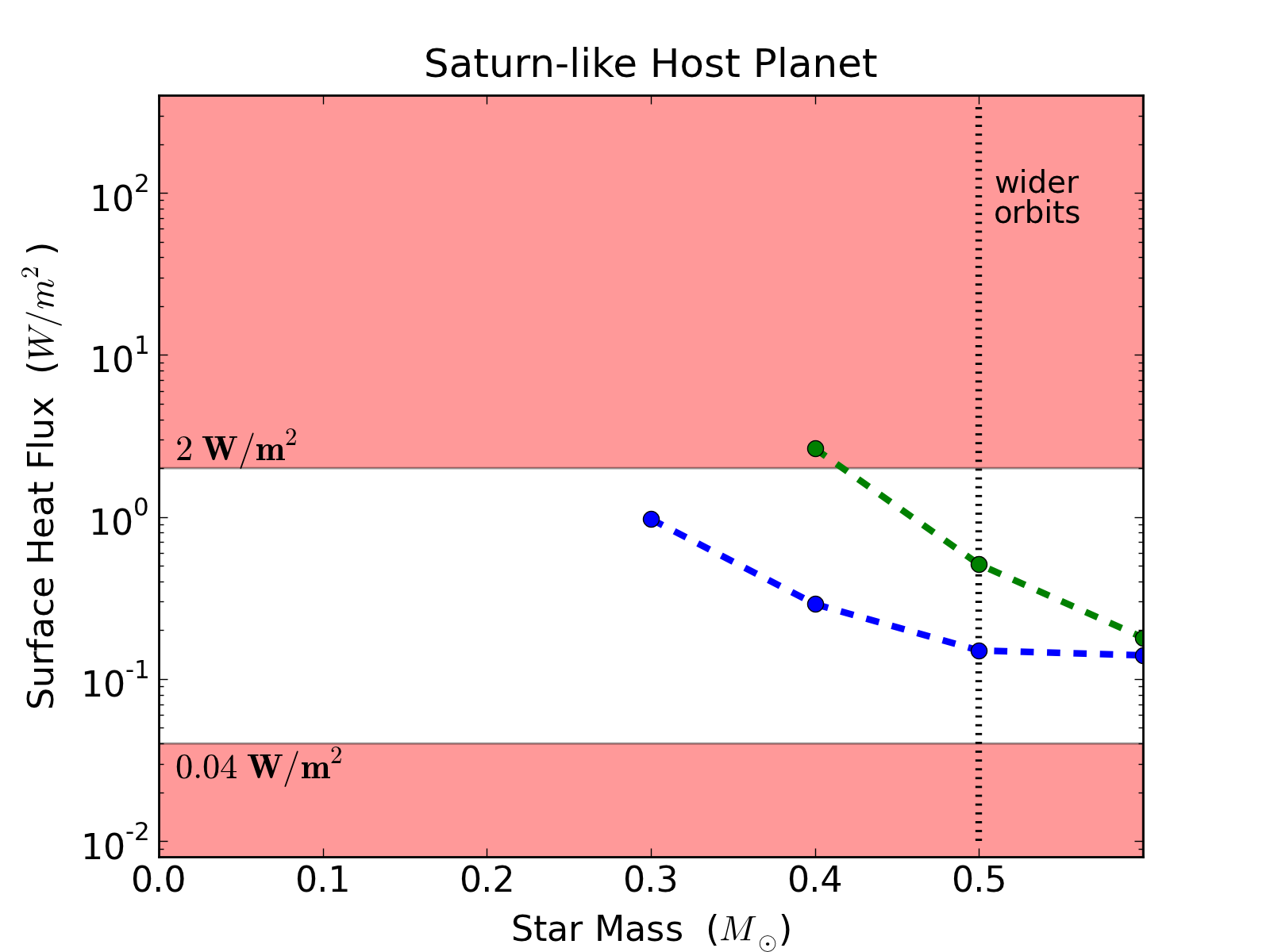}
    	   \end{array}$
     \includegraphics[width=.8\textwidth]{fig89legend.png}
  \caption{Graphical representations of the surface heat fluxes listed in Tables \ref{tab:cenend} and \ref{tab:eqend}. The red shaded regions represent exomoon tidal heating rates above $h_{\rm max}\equiv2\,{\rm W\,m}^{-2}$ or below $h_{\rm min}\equiv0.04\,{\rm W\,m}^{-2}$. The dashed vertical lines at 0.4 and 0.5 $M_{\odot}$ are where Ganymede-like orbits (wider orbits) begin to be gravitationally stable towards higher stellar masses.}
   \label{fig:b3ends}
\end{figure*}

It is important to compare the 3-body results with the provided 2-body result (the shaded rows) for each moon distance in Tables \ref{tab:cenend} and \ref{tab:eqend}. In each case, the difference in tidal heating between the 3-body and 2-body models decreases with increasing star mass. In other words, the influence of the star can be seen to decrease as the HZ moves outward for higher mass stars. Other than these general similarities, we will discuss each table separately. A graphical representation of the surface heat flux in each table, as a function of stellar mass, is also provided in Fig.~\ref{fig:b3ends}.\\*
\\*
\underline{Satellites Systems at the Center of the HZ (Table \ref{tab:cenend})}

In the tables, surface heat flux values highlighted in red are above $h_{\rm max}=2\,{\rm W\,m}^{-2}$ and represent a tidally evolved, steady state value from the 3-body simulations. Continuing with the assumption that exomoon habitability might be in jeopardy above $h_{\rm max}$, the table indicates that most of the moons would have the potential for habitability.  

The highest surface heat flux involves an Io-like orbit around a Jupiter-like planet with a 0.1 solar mass central star. This was the only stable satellite orbit for a 0.1 $M_{\odot}$ star and represents the tightest 3-body simulation (smaller star systems were considered, but none were stable). The effects of such a tight system can be seen by the long-term excitation of the moon in comparison to an isolated planet-moon binary whose orbit was completely circularized during the same time period. It is interesting that the final eccentricity is not unreasonably large ($e$ = 0.035), yet it is more than enough to generate extreme heating when combined with the short orbital distance. Heating rates for star masses of 0.2 $M_{\odot}$ are significantly lower in comparison. However, the surface fluxes are still about 3 times more than the most geologically active surface in our solar system, so we will exclude them from habitability considerations as well. Notice that all systems with a 0.2 $M_{\odot}$ star had heating rates well above our maximum for habitability.

There were mixed results for a star mass of 0.3 $M_{\odot}$. Both planet models had stable Europa-like moon orbits and the moons were comfortably below $h_{\rm max}$. For Io-like moon orbits the surface heat flux was below $h_{\rm max}$ for our model, but just above that for the conventional model. Given the inherent uncertainty in the tidal heating estimates, rather than discussing which model gives the better estimate, it is more relevant at this point to discuss the global flux ($F_{\rm glob}$) observed. Excluding the previously mentioned case of extreme tidal heating, the global flux for all the satellites is $\approx110\,{\rm W\,m}^{-2}$. Although the global flux does not lead to a direct estimate for the exomoon's surface temperature, a useful comparison can be made with the critical flux ($F_{\rm RG}$) estimate of $269\,{\rm W\,m}^{-2}$ for a runaway greenhouse in a Mars-mass exomoon. Clearly, the habitability of exomoons at this location is not at risk based on runaway greenhouse conditions from the global energy flux. In this case, tidal heating may actually be beneficial in warming large bodies of surface water that would otherwise freeze, which might be helpful also beyond the stellar HZ \citep{1987AdSpR...7..125R,2014AsBio..14...50H}.

Central star masses of 0.4 $M_{\odot}$ mark a cutoff for all exomoon habitability concerns based on intense tidal heating alone, since all the hypothetical exomoons are comfortably below $h_{\rm max}$ at this point. The star does influence the short orbit moons, but the influence may be seen as a benefit in promoting surface activity in the exomoons rather than a restriction for habitability. 

For the tidally evolved short orbits there is not a large difference in the final satellite heating rates between the two planet models and a given star mass. For example, with a star mass of 0.3 $M_{\odot}$ and an Io-like moon orbit, the moon around Jupiter had $h=1.49\,{\rm W\,m}^{-2}$. In comparison, the moon around Saturn had $h=1.56\,{\rm W\,m}^{-2}$. Some examples match even closer. The same cannot be said for the orbital elements, yet, these parameters would have to vary in order to compensate for the unique physical characteristics of the planet.  Given the physical differences and the corresponding difference in orbital distances for the moons, we did not expect the tidally evolved heating rates to be so similar.
 
As explained earlier, the 3-body simulation results involving the center of the HZ helped determine our second location for exploration inside the zone. Considering that the majority of the satellites had heating rates below $h_{\rm max}$ and that the global flux was well below the critical flux for a runaway greenhouse, it was clear that the next round of simulations should involve the inner HZ. \\ \\*

\underline{Satellites Systems at Earth-equivalent Distances (Table \ref{tab:eqend})}

For the second round of 3-body simulations we essentially moved the planet-moon binary closer to the star in the prior 3-body systems. By decreasing the orbital distance from the star the Hill radius of the planet also decreased. This led to a reduction in stable satellite systems, especially for wider orbits.

All stable satellite systems involving a central star of mass 0.4 $M_{\odot}$ and below experience surface heating greater than $h_{\rm max}$. Since the surface heat fluxes represent tidally evolved, steady state values, the heating has the potential to be maintained for extended periods of time. There is only small differences in final heating rates between specific moon orbits around Jupiter or Saturn-like host planets.

In regards to global flux, extreme tidal heating caused one satellite to be above the critical flux of $269\,{\rm W\,m}^{-2}$. In that situation the extreme heating was already enough to rule it out for habitability. The rest of the satellites are comfortably below the critical flux, leaving the surface heat flux as our primary consideration for exomoon habitability in these systems. At least, that is the case for the chosen physical parameters. An interesting future study would be to vary parameters in regards to the global flux and then investigate changes to the tidal evolution of each system.

\section{Conclusions}

We performed a computational exploration into specific characteristics of putative exomoon systems. Our study focused on satellite systems in the habitable zone (HZ) of \textcolor{black}{low-mass} stars, which contain significantly reduced HZ distances in comparison to Sun-like stars. For our planet-moon binaries the close proximity of the star presents dynamical restrictions to the stability of the moon, forcing it to orbit close to the planet to remain gravitationally bound. At short orbital distances, tidal heating and tidal torques between the planet and moon become more substantial. The relatively close star can also influence the long-term tidal evolution of the moon by continually exciting its orbit through gravitational perturbations. Key results from the computational simulations are highlighted as follows:

\noindent$\bullet$ \textbf{M dwarfs with masses $\bm{\lesssim 0.2\ M_{\odot}}$ cannot host habitable exomoons within the stellar habitable zone. }

Considering planets up to one Jupiter mass, stars with masses $\lesssim 0.2\ M_{\odot}$ can support a Mars-like satellite out to distances of $\sim10\ R_{\rm p}$ (Europa-like orbits). Though these systems are gravitationally stable, the close proximity to the star continually excites non-zero eccentricities in the moon's orbit. With limited orbital distances, tidal heating in the moon was significant, resulting in surface heating rates well above the proposed limit of $h_{\rm max}=2\,{\rm W\,m}^{-2}$. Considering the excessively high heating estimates these systems are unlikely to host habitable exomoons. This conclusion is consistent with our theoretical estimate based on Eqs.~(\ref{ahz}) - (\ref{minstarmass}).

Our results confirm that exomoon habitability is more complex than traditional definitions for planet habitability, which are based primarily on irradiation from a host star. Massive moons in the stellar HZ are not necessarily habitable by definition. Since the intense heating rates in the hypothetical exomoons are maintained by stellar perturbations, moving the host planet slightly outside the HZ should reduce the stellar influence. In this case, tidal heating in exomoons beyond the stellar HZ could make up for the reduced stellar illumination so that adequate surface temperatures for liquid surface water could be maintained.  

\noindent$\bullet$ \textbf{In the mass range $\bm{0.2\ M_{\odot} < M_\star < 0.5\ M_{\odot}}$, perturbations from the central star may continue to influence the long-term tidal evolution of exomoons in the stellar habitable zone.}

For stellar masses $\gtrsim 0.3\ M_{\odot}$, the distinction for habitable exomoons became less defined when based primarily on tidal evolution. Our results suggest that the host planet's location in the HZ has to be taken into consideration. Results from simulations involving Earth-equivalent distances (the inner HZ) show that M dwarfs with masses $\lesssim 0.4\ M_{\odot}$ promote surface heating beyond our accepted limit for habitability. In comparison, planetary orbits in the center of the HZ are within the established limits for habitability. Since a star's influence on a moon decreases with distance, so does its ability to excite higher heating rates. Therefore, this result for the center of HZs can be applied, by extension, to outer HZs. 

Our adopted maximum limit for habitability ($h_{\rm max}=2\,{\rm W\,m}^{-2}$) is based on a single example -- Jupiter's volcanically active moon Io. Although there is little doubt that Io does not support a habitable environment, there is no evidence that a Mars-sized or even an Earth-sized exomoon could not remain habitable given the same internal heating rate. This is especially true considering the uncertainty in the exact mechanisms of tidal dissipation and the efficiency of plate tectonics in terrestrial bodies. For these reasons, $h_{\rm max}$ should be considered more as a unit of comparison than a hard cutoff for habitable exomoons. The ultimate constraint on surface habitability will be given by the runaway greenhouse limit. With this consideration and treating the HZ as a whole, perturbations from a central star may continue to have deleterious effect on exomoon habitability up to $\approx 0.5\ M_{\odot}$. 

In contrast to the perils of intense tidal heating, perturbations from the star may actually have a positive influence on the habitability of exomoons. Figure~\ref{fig:b3ends} shows many surface heating rates below $h_{\rm max}$, yet above the proposed minimum for habitability ($h_{\rm min}=0.04\,{\rm W\,m}^{-2}$). These results suggest that satellite systems around stars in this mass range would need to be considered on a case by case basis depending on the planet mass and the specific location in the HZ. Moderate tidal heating might actually help to sustain tectonic activity (thus possibly a carbon-silicate cycle) and an internal magnetic dynamo that might help to shield the moon from high-energy radiation \citep{hz13} on a Gyr timescale.

\noindent$\bullet$ \textbf{Considerations of global energy flux do not restrict habitability of exomoons in the HZs around stars with masses above $\bm{0.2\ M_{\odot}}$.}

The conclusions thus far are in agreement with predictions made by \cite{heller12} who followed considerations of energy flux and gravitational stability. This study, which focused on tidal evolution, has verified those predictions using a tidal model that considered both $N$-body interaction and tidal evolution. Similar considerations for energy flux were incorporated within this study with the global averaged flux ($F_{\rm glob}$) listed in Tables \ref{tab:cenend} and \ref{tab:eqend}. Compared to the critical flux of $269\,{\rm W\,m}^{-2}$ for a runaway greenhouse on a Mars-mass satellite,  the 3-body global flux results support the conclusion that star masses $\lesssim 0.2\ M_{\odot}$ are unlikely to host habitable exomoons. Above that mass, exomoon habitability was not constrained by global energy flux.

\noindent$\bullet$ \textbf{Torques due to spin and tidal distortion between the planet and moon can result in rapid inward spiraling of a moon for orbital distances $\bm{\lesssim 6\ R_{\rm p}$}.}

In specific simulations involving a Saturn-like host planet and Io-like ($a \sim6\ R_{\rm p}$) moon orbits, distortion torques resulted in the complete inward spiral of a moon in $<10^7$\,yr. The inward migration was connected to the assumption that the giant host planet was tidally locked to the star. While the orbital decay rate was slower for more massive Jupiter-like host planets, a conservative estimate for the maximum lifetime of Io-like moon orbits was only 200\,Myr. Compared to the geological age of the Earth, this is a short lifespan. Assuming a Mars-like moon in an Io-like orbit was initially habitable, implications for the development of life may be considerable.

\subsection{Future Work}

For the sake of minimizing computational demands we assumed coplanar orbits and no spin-orbit misalignments of the moons. However, we expect new effects to arise in more complex configurations, such as spin-orbit resonances between the circumstellar and circumplanetary orbits as well as substantial effects on the longterm evolution of tidal heating. Important observational predictions can be obtained by studying the obliquity evolution of initially misaligned planet spins due to tidal interaction with the star. Will this ``tilt erosion'' \citep{2011OLEB...41..539H} tend to align the moons' orbits with the circumstellar orbit? These investigations could help predicting and interpreting possible variations of the planetary transit impact parameter due to the presence of an exomoon \citep{kipping09b}.

Of specific interest are retrograde satellites, which we neglected in this study. These satellites can form through direct capture or tidal disruption during planet-planet encounters \citep{2006Natur.441..192A,2013AsBio..13..315W}, and they can be orbitally stable as far as the Hill radius \citep{domingos06}. Hence, we expect that retrograde moons could still be present or habitable in some cases that we identified as unstable or even uninhabitable for prograde moons. What is more, the detection of an exomoon in a very wide circumplanetary orbit near or beyond about 0.5 Hill radii might only be explained by a retrograde moon. 

Owing to computational restrictions, \textcolor{black}{the number of bodies considered was limited to three. A study with additional bodies would be of interest. The effects of orbital resonances for mutliple moons could be considered as well as the influence of additional planets in the system just outside the HZ.} It was also necessary to limit many physical parameters such as the moments of inertia, tidal dissipation factors ($Q$), and tidal Love numbers ($k_{L}$). More fundamentally, these parameters are treated as constants in our applied tidal model. However, under the effect of tidal heating, the rheology and tidal response of a moon (or planet) can change substantially \citep{2014ApJ...789...30H,2015ApJ...804...41D}. Hence, more advanced tidal theory needs to be incorporated in our mathematical treatment of tidal plus $N$-body interaction to realistically model tidal effects in the regime of enhanced tidal heating. \textcolor{black}{Modifying these values can coincide with variations in other physical parameters such as mass and radius for the extended bodies, which should noticeably affect the dynamical stability and tidal evolution of the systems.}

One result from this study is the hypothetical existence of extremely tidally heated moons. \cite{peters13} proposed the direct imaging of tidally heated exomoons. Closer examination is warranted to see if our computer model would be useful in providing orbital constraints on directly detectable exomoons. When considering extreme heating in a massive body, the issue of inflation may become important. Inflation is a physical response that was not incorporated into our model. Planetary inflation was considered by \cite{mardling02} and future plans include the integration of this effect into our tidal evolution code.

\section{Acknowledgements}
Rhett Zollinger was supported by funding from the Utah Space Grant Consortium. Ren{\'e} Heller has been supported by the Origins Institute at McMaster University, by the Canadian Astrobiology Program (a Collaborative Research and Training Experience Program funded by the Natural Sciences and Engineering Research Council of Canada), by the Institute for Astrophysics G\"ottingen, by a Fellowship of the German Academic Exchange Service (DAAD), and by the German space agency (Deutsches Zentrum f\"ur Luft- und Raumfahrt) under PLATO grant 50OO1501. This work made use of NASA's ADS Bibliographic Services. We would like to thank Benjamin C. Bromley, from the University of Utah, for his valuable insight and advice regarding this work. Finally, we would like to thank the reviewers for their helpful suggestions and useful insights.

\bibliographystyle{mnras}
\bibliography{exomoonRefs}

\bsp	
\label{lastpage}
\end{document}